\documentclass[journal]{IEEEtran}
\usepackage{xcolor}
\usepackage{tikz}
    \usetikzlibrary{positioning}
    \usetikzlibrary{calc}
    \usetikzlibrary{shapes}
    \usetikzlibrary{plotmarks}
    \usepackage{tikz-qtree}
    \usepackage[edges]{forest}
    \usetikzlibrary{arrows.meta}
    \usetikzlibrary{patterns}
\usepackage{pgfplots}
    \pgfplotsset{compat=newest}
\usepackage{standalone}
\usepackage{mathtools}
\usepackage{dsfont}
\usepackage{wasysym}
\usepackage{amssymb}
\usepackage{amsfonts}
\usepackage{amsthm}
    \theoremstyle{definition}
    \newtheorem{example}{Example}
    \newtheorem{remark}{Remark}
    \theoremstyle{definition}

\usepackage[hidelinks]{hyperref} 
    \hypersetup{final} 
\usepackage{cite}

\colorlet{UniformColor}{red!70!black}%
\colorlet{ESS_Color}{green!70!black}%
\colorlet{SM_Color}{brown!70!black}%
\colorlet{MPDM_Color}{blue!70!black}%
\colorlet{CCDM_Color}{orange!70!black}%

\pgfplotsset{Uniform/.style={color=UniformColor,mark=square*,mark size=1.8pt, thick,mark options={fill=white,solid}, thick}}
\pgfplotsset{Uniform2/.style={color=UniformColor, thick}}
\pgfplotsset{Uniform3/.style={color=UniformColor, very thick}}
\pgfplotsset{CCDM/.style={color=CCDM_Color,mark=pentagon*,mark size=1.8pt, thick,mark options={fill=white,solid}, thick}}
\pgfplotsset{CCDM2/.style={color=CCDM_Color,thick}}
\pgfplotsset{MPDM/.style={color=MPDM_Color,mark=diamond*,mark size=1.8pt, thick,mark options={fill=white,solid}, thick}}
\pgfplotsset{MPDM2/.style={color=MPDM_Color,thick,dashed}}
\pgfplotsset{ESS/.style={color=ESS_Color,mark=*,mark size=1.8pt, thick,mark options={fill=white,solid}, thick}}
\pgfplotsset{ESS2/.style={color=ESS_Color, thick}}
\pgfplotsset{SM/.style={color=SM_Color,mark=x,mark size=1.8pt, thick,mark options={fill=white,solid}, thick}}
\pgfplotsset{Capacity/.style={color=black,thick}}
\pgfplotsset{Capacity2/.style={color=black,very thick}}
\def\calA{\mathcal{A}}

\def\calS{\mathcal{S}}

\def\calX{\mathcal{X}}
\def\calY{\mathcal{Y}}

\def\boldP{\boldsymbol{P}}

\def\rloss{R_{\text{loss}}}

\def\spsh{SpSh}

\def\Rs{R_{\text{s}}}

\def\snr{\text{SNR}}

\def\exp{\mathbb{E}}
\def\ent{\mathbb{H}}

\def\rbmd{R_{\text{BMD}}}
\def\capc{C}
\def\delsnr{\Delta\text{SNR}}

\colorlet{shapercolor}{green!20!white}
\colorlet{bshapercolor}{cyan!20!white}
\colorlet{feccolor}{blue!20!white}
\colorlet{mapcolor}{red!20!white}
\colorlet{chancolor}{black!10!white}

\def\lc{\left\lceil}   
\def\rc{\right\rceil}
\def\lf{\left\lfloor}   
\def\rf{\right\rfloor}

\def\calAspd{\calA^\star}

\def\N{n}
\def\Comp{C}
\let\floor\relax
\DeclarePairedDelimiter\floor{\lfloor}{\rfloor}

\newcommand{\MultinomCoeff}[1]{\ensuremath{\textrm{MC}\!\left( #1 \right)}}

\def\namp{n_a}

\begin{document}
\title{Probabilistic Shaping for Finite Blocklengths: Distribution Matching and Sphere Shaping}

\author{Yunus~Can~G\"{u}ltekin,~\IEEEmembership{Student~Member,~IEEE,}
Tobias~Fehenberger,~\IEEEmembership{Member,~IEEE,}
Alex~Alvarado,~\IEEEmembership{Senior~Member,~IEEE,}
and~Frans~M.~J.~Willems,~\IEEEmembership{Fellow,~IEEE}
\thanks{Y. C. G\"{u}ltekin, A. Alvarado and  F. M. J. Willems are with the Information and Communication Theory Lab, Signal Processing Systems Group, Department of Electrical Engineering, Eindhoven University of Technology, Eindhoven 5600 MB, the Netherlands (e-mails: \{y.c.g.gultekin, a.alvarado, f.m.j.willems\}@tue.nl).}
\thanks{T. Fehenberger was with the Information and Communication Theory Lab, Signal Processing Systems Group, Department of Electrical Engineering, Eindhoven University of Technology, Eindhoven 5600 MB, the Netherlands. He is now with ADVA Optical Networking, Munich 82152, Germany (e-mail: tfehenberger@advaoptical.com).}
\thanks{The work of Y. C. G\"{u}ltekin is supported by TU/e Impuls program, a strategic cooperation between NXP Semiconductors and Eindhoven University of Technology.
The work of A. Alvarado is supported by the Netherlands Organisation for Scientific Research (NWO) via the VIDI Grant ICONIC (project number 15685) and has received funding from the European Research Council (ERC) under the European Union's Horizon 2020 research and innovation programme (grant agreement No 57791).}}

\markboth{Preprint, \today}{}

\maketitle

\begin{abstract}
In this paper, we provide for the first time a systematic comparison of distribution matching (DM) and sphere shaping (\spsh) algorithms for short blocklength probabilistic amplitude shaping. 
For asymptotically large blocklengths, constant composition distribution matching (CCDM) is known to generate the target capacity-achieving distribution. 
As the blocklength decreases, however, the resulting rate loss diminishes the efficiency of CCDM. 
We claim that for such short blocklengths and over the additive white Gaussian channel (AWGN), the objective of shaping should be reformulated as obtaining the most energy-efficient signal space for a given rate (rather than matching distributions). 
In light of this interpretation, multiset-partition DM (MPDM), enumerative sphere shaping (ESS) and shell mapping (SM), are reviewed as energy-efficient shaping techniques. 
Numerical results show that MPDM and \spsh~have smaller rate losses than CCDM. 
\spsh---whose sole objective is to maximize the energy efficiency---is shown to have the minimum rate loss amongst all. 
We provide simulation results of the end-to-end decoding performance showing that up to 1 dB improvement in power efficiency over uniform signaling can be obtained with MPDM and \spsh~at blocklengths around 200. 
Finally, we present a discussion on the complexity of these algorithms from the perspective of latency, storage and computations.
\end{abstract}

\IEEEpeerreviewmaketitle

\section{Introduction}\label{sec:intro}
Coded modulation (CM), which combines multi-level modulation with forward error correction (FEC), is indispensable for digital communication strategies targeting high transmission rates.
To realize CM, different techniques have been proposed in the literature, such as multilevel coding (MLC)~\cite{Imai1977_TransIT_multilevel,wachsmann1999}, trellis CM~\cite{Ungerbock1982_TransIT_TCM}, and bit-interleaved CM (BICM)~\cite{Zehavi1992_TCOM_bicm,Caire1998_TransIT_BICM,Albert2008_BICM,Martinez2009_TransIT_bicm,Szczecinski2015_BICMbook}.
Among the many proposed CM architectures, the de-facto standard is to combine high-order modulation format with a binary FEC code using a binary labeling strategy, frequently in the absence of an interleaver, and to use bit-metric decoding (BMD) at the receiver~\cite{Martinez2009_TransIT_bicm}, which corresponds to the BICM paradigm.

As the modulation order increases, the maximum rate that can be achieved with uniform signaling starts to suffer from a loss with respect to the channel capacity\footnote{The channels under consideration here are assumed to have nonuniform capacity-achieving distributions.}.
As an example, the maximum achievable information rate (AIR) for MLC in combination with multi-stage decoding (MSD)~\cite{wachsmann1999} is the mutual information (MI) of the channel input and output.
If a uniform signaling strategy is employed with MLC-MSD, the MI is bounded away from capacity.
This gap is called the \emph{shaping gap} and is up to 0.255 bits per real channel use (bit/1-D) for the additive white Gaussian noise (AWGN) channel.
When translated into an increase in required signal-to-noise ratio (SNR) to obtain a certain MI, this so-called \emph{ultimate shaping gap} corresponds to a 1.53~dB loss in energy efficiency~\cite{forney1984}.
The well-known 1.53~dB is an asymptotic result for the AWGN channel and is only relevant for CM systems where the maximum AIR is the MI, and when the number of channel uses $n$ as well as the modulation order approach infinity. 

\begin{figure}[t]
\centering		
\includegraphics[width=\columnwidth]{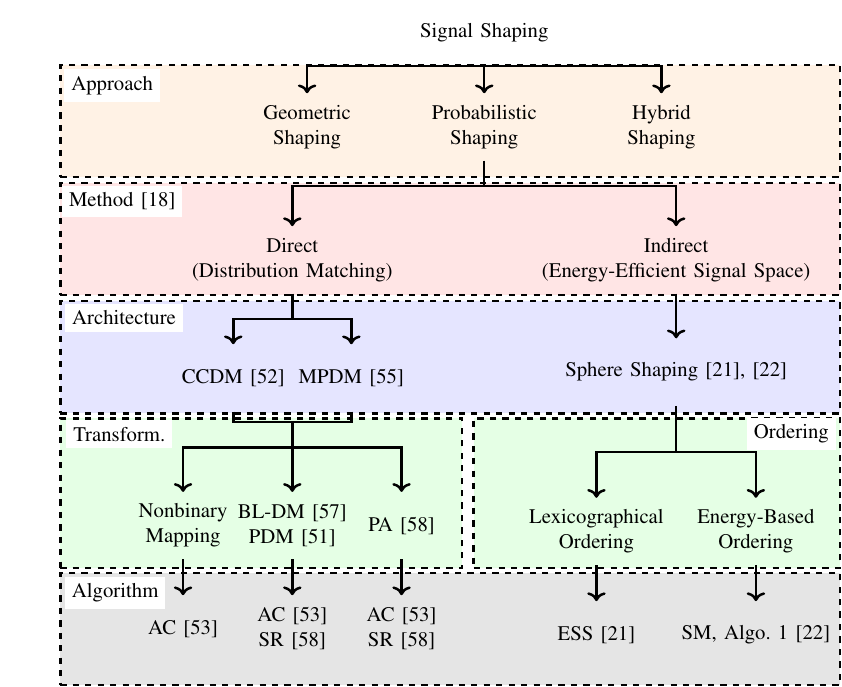}
\caption{Taxonomy of shaping in the context of PAS. We focus on the schemes that are evaluated in this work.}
\label{fig:shapingtaxonomy}
\end{figure}

There exist numerous techniques in the literature, most of them proposed in the late 1980s and early 1990s, that attempt to close the shaping gap.
Motivated by the fact that the capacity-achieving distribution for the AWGN channel is Gaussian, these techniques fundamentally aim at one of the following.
The first goal is to construct a signal constellation with a Gaussian-like geometry, which is called geometric shaping (GS)~\cite{Sun1993_ITtrans_geometricshap,LoghinZMAKP2016_NonUnifConstforATSC,QuD2017_GSoutperformPS,SteinerB2017_CompareGSPS,Boutros2018_geometricshap,ChenOHA2018_IncreaseAIRviaGS,ChenOLA2018_GS64viaAIR,ChenLLOA2018_RateAdaptviaGS}.
The other approach is to induce a Gaussian-like distribution over the signal structure, which is called probabilistic shaping (PS)~\cite{calderbank1990,Forney1992_ITtrans_trellisshap,kschischang1993,willems1993,laroia1994}.
PS techniques can be further classified into two subgroups using the terminology introduced by Calderbank and Ozarow in~\cite{calderbank1990}.
The \emph{direct} approach is to start with a target distribution (which is typically close to the capacity-achieving distribution) on a low-dimensional signal structure and have an algorithm try to obtain it~\cite{calderbank1990,kschischang1993}.
Following recent literature~\cite{BochererM2011_MatchingDistributions}, the direct approach can also be called distribution matching (DM).
The \emph{indirect} approach is to start with a target rate and bound the multi-dimensional signal structure by a sphere, which we call sphere shaping (\spsh)~\cite{willems1993,laroia1994}. 
Here, a Gaussian distribution is induced indirectly (when $n\to\infty$) as a by-product.
Finally, there exist some \emph{hybrid shaping} approaches in which GS and PS are combined~\cite{Batshon2017_ecoc_hybrid,Cai2018_JLT_hybridshaping,Cai_JLT_hybridshaping_2}.
We refer to~\cite[Sec. 4.5]{Fischer2002_PrecodingShaping} for a detailed discussion on GS, and to~\cite[Ch. 4]{Fischer2002_PrecodingShaping} and~\cite[Sec. II]{bocherer2015} on PS.
GS, PS, and hybrid shaping are shown on the top layer of Fig.~\ref{fig:shapingtaxonomy} where the taxonomy of constellation shaping (as discussed in the current paper) is illustrated.
We call this first layer {\it shaping approach}.
On the second layer which we call {\it shaping method}, PS is split into two following the Calderbank/Ozarow terminology~\cite{calderbank1990}.

In the context of BICM, signal shaping techniques again attracted a considerable amount of attention in the 2000s.
GS was investigated for BICM in~\cite{SommerF2000_ShapingforQAMAWGNTurbo,LeGoff2003_Constellations4BICM,BarsoumJF2007_ConstDesignviaCmax}, and PS was studied in~\cite{LeGoffSJ2004_BICMwithShaping,RaphaeliG2004_ShapingforTurboCM,LeGoffSJ2005_BICMwithShaping_comlett,ValentiZ2012_ShapingforBICMLDPCAPSK}.
An iterative demapping and decoding architecture with PS was proposed in~\cite{LeGoffKTS2007_ShapingTurboCMiterative}.
The achievability of the so-called generalized MI (GMI) was shown for independent but arbitrarily distributed bit-levels in~\cite{FabregasM2010_BICMwShaping}.
In~\cite{AlvaradoBA2011_HighSNRforBICMcapacity}, it was demonstrated that the GMI is a nonconvex function of the input bit distribution, i.e., the problem of computing the input distribution that maximizes GMI is nonconvex.
An efficient numerical algorithm to compute optimal input distributions in BICM was introduced in~\cite{BochererAACM2012_EfficientAlgoBICMcapacity}.
The effect of mismatched shaping, i.e., not using the true symbol probabilities or reference constellation at the receiver, was examined in~\cite{PengFM2012_MismatchedShaping}.
The achievable rates, error exponents and error probability of BICM with PS were analyzed in~\cite{Peng2012PhdThesis}.
Signal shaping was investigated for BICM at low SNR in~\cite{AgrellA2013_SignalShapingforBICMatlowsnr}.
PS in BICM was considered for Rayleigh fading channels in~\cite{BouazzaD2007_BICMiterativeShapingFading,XiangV2011_DVBS2shapingiterative}.

\begin{figure*}[t]
\centering		
\includegraphics[width=\columnwidth]{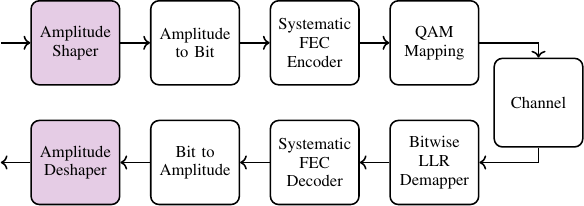}
\hspace{5mm}
\includegraphics[width=0.9\columnwidth]{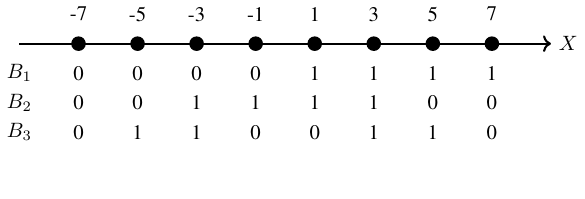}
\caption{(Left) Block diagram of the PAS architecture. Amplitude shaping blocks (green boxes) are examined in the current paper. (Right) The binary reflected Gray code (BRGC) for 8-ASK. A quadrature amplitude modulation (QAM) symbol is the concatenation of two ASK symbols.}
\label{fig:PAShighlevel}
\end{figure*}

Recently, probabilistic amplitude shaping (PAS) has been proposed to provide low-complexity integration of shaping into existing binary FEC systems with BMD~\cite{bocherer2015}.
PAS uses a reverse concatenation strategy where the shaping operation precedes FEC coding, as shown in Fig.~\ref{fig:PAShighlevel} (left).
This construction has been first examined for constrained coding problems~\cite{bliss1981}. 
A corresponding soft-decision decoding approach for this structure was studied in \cite{fan1999}.
PAS can be considered as an instance of the Bliss architecture~\cite{bliss1981} where in the outer layer a shaping code is used, and then in the inner layer parity symbols are added.
The main advantage of this structure is that amplitude shaping can be added to existing CM systems as an outer code.
In addition to closing the shaping gap, PAS moves the rate adaptation functionality to the shaping layer.
This means that, instead of using many FEC codes of different rates to obtain a granular set of transmission rates, the rate is adjusted by the amplitude shaper with a fixed FEC code.
Owing to these advantages, PAS has attracted a lot of attention. PAS has been combined with low-density parity-check (LDPC) codes~\cite{bocherer2015}, polar codes~\cite{polarccdm} and convolutional codes~\cite{Gultekin2019Arxiv_ESS}. 
Its performance has been evaluated over AWGN channels~\cite{bocherer2015}, optical channels~\cite{Buchali2016_jlt,fehenberger2016}, wireless channels~\cite{Gultekin2019Arxiv_ESS} and parallel channels with channel state information available at the transmitter~\cite{SteinerSB2018_PDM}. 

The key building blocks of the PAS framework are the amplitude shaper and deshaper, i.e., the purple boxes in Fig.~\ref{fig:PAShighlevel} (left). 
The function of the amplitude shaper is to map uniform binary sequences to shaped amplitude sequences in an invertible manner.
A careful selection of the set of sequences that can be outputted by the shaper with the aim of matching a target distribution (direct approach) or constructing an energy-efficient signal space (indirect approach) results in improvement in overall performance.
We call the way this selection is accomplished {\it shaping architecture} which affects the performance of PAS.
On the other hand, the actual implementation of this architecture is called here the {\it shaping algorithm} and determines the complexity of attaining this performance.
The third and fifth layers of Fig.~\ref{fig:shapingtaxonomy} illustrate  shaping architectures and algorithms, respectively.
The difference between the shaping architecture and the underlying algorithm is discussed in detail in~Sec.~\ref{ssec:dm_architecture_vs_algorithm}.

For the initial proposal of PAS~\cite{bocherer2015}, constant composition distribution matching (CCDM) was employed as the shaping architecture~\cite{ccdm}. 
The basic principle of CCDM is to utilize amplitude sequences having a fixed empirical distribution that is information-theoretically close to the target distribution. 
To this end, a constant composition constraint is put on the output sequences such that all have the same amplitude composition.
To realize such a mapping, arithmetic coding (AC) is used in a way similar to~\cite{ramabadran1990}.
Although CCDM has vanishing rate loss for asymptotically large blocklengths~\cite{ccdm}, it has two fundamental drawbacks that prohibit its use for finite blocklengths.
First, as recently shown in~\cite{GultekinHW2018_OnShaping4ShortBlocks} and~\cite[Fig. 4]{Fehenberger2019TCOM_MPDM}, CCDM suffers from high rate losses as the blocklength decreases.
Second, CCDM is implemented based on AC which is strictly sequential in input length~\cite{ramabadran1990},\cite[Ch. 5]{sayood2002lossless}.

To replace CCDM in the short-to-moderate blocklength regime and to provide more hardware-friendly implementations, improved techniques have been devised.
The most prominent DM examples other than CCDM include multiset-partition DM (MPDM)~\cite{Fehenberger2019TCOM_MPDM} and product DM (PDM)~\cite{SteinerSB2018_PDM,pikus2017}.
Briefly stated, MPDM uses different compositions and expands the set of output sequences to achieve smaller rate losses than CCDM.
With the same objective, PDM internally uses multiple binary matchers to generate the desired distribution as a product distribution\footnote{A symbol-level product distribution can be written as the product of bit-level distributions~\cite[eq. (14)]{SteinerSB2018_PDM}. In the context of BICM, product distributions were studied extensively in~\cite{Peng2012PhdThesis}.}.
In~\cite{Fehenberger2019Arxiv_PASR}, a parallel-amplitude (PA) architecture is proposed for DM to enable even higher degree of parallelization.
Also in~\cite{Fehenberger2019Arxiv_PASR}, subset ranking (SR) is introduced as an alternative to the conventional AC method for binary-output CCDM.
As for direct shaping methods, enumerative sphere shaping (ESS) and shell mapping (SM) are notable \spsh~algorithms which are initially proposed in~\cite{willems1993} and~\cite{LangL1989_LeechLatticeModem}, respectively.
ESS is recently considered in PAS framework~\cite{GultekinHSW2017_ESSfor80211,Gultekin2019Arxiv_ESS,Amari2019_IntroducingESSoptics,Amari2019_ESSreachincrease,Goossens2019_FirstExperimentESS}, as well as SM in~\cite{Schulte2019_commlett_shellmapp}.
Furthermore, low-complexity implementation ideas for both of these algorithms have been presented in~\cite{GultekinWillems2018ISIT_EnumerativeShaping}.

The fourth layer in Fig.~\ref{fig:shapingtaxonomy} which we call {\it transformation} for DM and {\it ordering} for \spsh~designates the way a shaping algorithm formulates a solution to the problem defined by the shaping architecture. 
As an example, CCDM considers sequences having the same composition~\cite{ccdm}. 
By realizing a binary-to-nonbinary transformation with AC~\cite{ramabadran1990,ccdm}, CCDM can directly be used to produce amplitude sequences. 
On the other hand, separate binary-to-binary transformations can be employed for different bit-levels using AC~\cite{ramabadran1990} or SR~\cite{Fehenberger2019Arxiv_PASR}.
Then these bit-levels can be combined such that the corresponding channel input distribution is close to the capacity-achieving distribution~\cite{pikus2017,SteinerSB2018_PDM}.
As another example, \spsh~considers amplitude sequences in a sphere.
ESS orders these sequences lexicographically~\cite{willems1993}, while SM and~\cite[Algorithm 1]{laroia1994} order them based on their energy.

Other shaping schemes have been proposed that are briefly listed in the following. 
A detailed analysis of them is outside the scope of this manuscript.
The concept of a ``mark ratio controller" was proposed for low-complexity implementation of BL-DM in~\cite{YoshidaKA2017_MarkRatioDM,YoshidaKA2018_PeriodicalUniformaliz}.
In the streaming DM of~\cite{streamingdm} and the prefix-free code distribution matching with framing of~\cite{ChoW2019_MultiRatePFDM,Cho2019_PrefFreeDM}, switching is performed between two (or more) variable-length shaping codes such that the output is always of fixed length.
In~\cite{PikusX2019_ACbasedMCBLDM}, a ``multi-composition" idea similar to~\cite{Fehenberger2019TCOM_MPDM} was applied to BL-DM.
The authors of~\cite{PikusXK2019_FPACDM} provided a finite-precision implementation for AC-CCDM.
In~\cite{Gultekin2019Arxiv_PESS}, a shaper based on ESS was introduced to shape a subset of the amplitude bit-levels, which is referred to as partial ESS (P-ESS).
The authors of~\cite{YoshidaKA2019_HIDM} introduced the ``hierarchical" DM which realizes a nonuniform distribution with hierarchical LUTs.
An approximate sphere shaping implementation based on Huffman codes was proposed in~\cite{Millar2019ECOC_HCSS}.

In this work, we examine DM and \spsh~methods.
The contributions of this paper are threefold.
First, this paper is---to the best of our knowledge---the first study where a systematic comparison of different PS architectures is provided.
Second, using rate loss as well as AIRs for finite-length shapers as the performance metrics, we claim that shaping strategies which aim to construct energy-efficient signal sets are more effective than the techniques which focus on matching distributions for the AWGN channel. 
For the analyzed schemes, this means that MPDM and \spsh, are more efficient for short blocklengths than CCDM whose sole objective is to obtain the capacity-achieving distribution.
Our claim is then verified via frame error rates (FERs) that are obtained in end-to-end decoding simulations of the PAS system employing long and short systematic LDPC codes from~\cite{DVBS2} and~\cite{80211-2016}, respectively.
The improvements in power efficiency that we obtained during end-to-end decoding simulations are consistent with the predictions made by finite-length AIRs.
The third contribution of this paper is to provide a discussion on the required storage, computational complexity, and the latency of different DM and \spsh~algorithms.

The paper is organized as follows.
The first part is tutorial-like.
In Sec.~\ref{sec:prelim}, background information on uniform and shaped signaling schemes, and amplitude shaping is provided.
Section~\ref{sec:shaping_schemes} reviews DM and \spsh~schemes from shaping architecture and algorithmic implementation perspectives.
The second part of the paper is reserved for the comparison of four amplitude shaping architectures.
Rate losses, AIRs, and end-to-end decoding performance of PAS are studied in Sec.~\ref{sec:performance_comparison}.
Section~\ref{sec:complexity} is devoted to a high-level discussion on latency and complexity of the schemes under consideration. 
Finally, conclusions are given in Sec.~\ref{sec:conclusion}.

\section{Preliminaries}\label{sec:prelim}
\subsection{Notation and Definitions}
We use capital letters $X$ to denote random variables, lower case letters $x$ to specify their realizations.
Random vectors of length $n$ are indicated by $X^n$ while their realizations are notated by $x^n$.
Element-wise multiplication of $x^n$ and $y^n$ is shown by $x^ny^n$.
Calligraphic letters $\calX$ represent sets.
The Cartesian product of $\calX$ and $\calY$ is indicated as $\calX \times \calY$, while $\calX^n$ is the $n$-fold Cartesian product of $\calX$ with itself.
Boldface capital letters $\boldP$ specify matrices.
Probability distributions over $\calX$ are denoted by $P_X(x)$.
The probability density function (PDF) of $Y$ conditioned on $X$ is indicated by $f_{Y|X}(y|x)$.

The discrete-time AWGN channel output is given at time $i = 1, 2,\cdots, n$ by $Y_i = X_i + Z_i$, where $Z_i$ is the noise which is independent of the input $X_i$, and drawn from a zero-mean Gaussian distribution with variance $\sigma^2$.
There is an average power constraint $\exp[X^2]\leq P$, where $\exp[\cdot]$ is the expectation operator.
The SNR is $\snr=\exp[X^2]/\sigma^2$.

The capacity of the AWGN channel is $\capc = \frac{1}{2}\log_2(1+\snr)$ in bit/1-D.
This capacity can be achieved as $n\rightarrow\infty$ by employing a codebook (set of input sequences) in which all the codewords (input sequences) are generated with entries independent and identically distributed according to a zero-mean Gaussian with variance $P$~\cite[Ch. 9]{CoverT2006_ElementsofInfoTheo}.
The corresponding random coding argument shows that channel input sequences, drawn from a Gaussian distribution, are likely to lie in an $n$-sphere of squared radius $n P(1+\varepsilon)$ for any $\varepsilon > 0$, when $n\rightarrow\infty$.
This motivates to select the signal points from within an $n$-sphere, or equivalently to use an $n$-sphere as the signal space boundary, in order to achieve capacity. 
For a more detailed discussion on the asymptotic duality of Gaussian distributions and $n$-spherical signal spaces for large $n$, we refer the reader to, e.g.,~\cite[Sec.~IV-B]{forney1984}

\subsection{Discrete Constellations and Amplitude Shaping}
We consider $2^m$-ary amplitude-shift keying (ASK) alphabets $\calX = \{ \pm 1, \pm 3,\cdots, \pm(2^m-1) \}$, which can be factorized as $\calX = \calS \times \calA$.
Here $\calS = \{\pm 1\}$ and $\calA = \{ 1, 3,\cdots, 2^{m}-1\}$ are the amplitude and sign alphabets, respectively.
The cardinality of the amplitude alphabet is $\namp = |\calA|$.
Motivated by the fact that the capacity-achieving distribution for the AWGN channel is symmetric around the origin, we restrict our attention to the amplitude distribution $P_A(a)$, and assume that the sign distribution $P_S(s)$ is uniform and independent of the amplitudes.
The distribution of the channel input $X = SA$ is then $P_X(x) = P_S(s)P_A(a)$.

The distribution that maximizes the MI for ASK constellations subject to an average power constraint does not have a known analytical form.
Maxwell-Boltzmann (MB) distributions $P_A(a) = K\left(\lambda\right) e^{-\lambda a^2}$ for $a \in \calA$,
are used for shaping amplitudes, e.g., in~\cite{kschischang1993,bocherer2015}, since they are the discrete-domain counterpart of the Gaussian distribution and maximize the entropy for a given average energy~\cite{CoverT2006_ElementsofInfoTheo}.
Furthermore, as shown in ~\cite[Table 5.1]{Bocherer2018Habilitation}, the difference in MI for the MB distribution and the capacity-achieving distribution is insignificant for ASK constellations.
For MB distributions, $\lambda$ determines the variance of the distribution while $K(\lambda)$ normalizes it.

In a dual manner, \spsh~is also employed for amplitude shaping in the discrete domain~\cite{willems1993,laroia1994}.
In~\cite{GultekinHW2018_OnShaping4ShortBlocks}, it is shown that when an $n$-spherical region of $\calX^n$ is used as the signal space, the distribution induced on $\calA$ approaches an MB distribution as $n\rightarrow\infty$.
The authors of~\cite{Schulte2019_commlett_shellmapp} showed that at finite $n$, \spsh~minimizes the informational divergence between the induced distribution and an MB distribution.

To employ high-order modulation formats such as $2^m$-ASK for $m\geq 2$, a binary labeling strategy is necessary.
A discussion on binary labeling can be found in~\cite[Sec. 2]{Szczecinski2015_BICMbook}.
We assume that the binary label $B_1B_2\cdots B_m$ of a channel input $X$ can be decomposed into a sign bit $B_1$ and amplitude bits $B_2B_3\cdots B_m$.

\begin{example}[{\bf Binary labeling}]\label{ex:BinaryLabeling}
The BRGC is tabulated for 8-ASK in Fig.~\ref{fig:PAShighlevel} (right).
Here, $B_1$ is symmetric around zero.
Furthermore, when $X$ has a distribution which is symmetric around zero, $B_1$ is uniform and stochastically independent of $B_2$ and $B_3$.
In this paper, we assume that the BRGC is used for labeling.
\end{example}

\subsection{Fundamentals of Amplitude Shaping Schemes} 
The amplitude shaper is a block that maps $k$-bit uniform sequences to $n$-amplitude shaped sequences in an invertible manner.
The tasks of this block are (i) to create a {\it shaping codebook} $\calAspd \subseteq \calA^n$, and (ii) to realize a {\it shaping encoder} to index these sequences. 
The former task is related to the properties of the desired set $\calAspd$ while the latter deals with the algorithmic implementation of the mapping.
This difference is discussed in detail in Sec.~\ref{ssec:dm_architecture_vs_algorithm}.
In the remainder of this section, we introduce the concepts and parameters that are associated with the shaping techniques that will be investigated in this paper. 

The energy of a sequence $a^n = a_1, a_2,\cdots, a_n$ is
\begin{align}
e\left(a^n\right) = \sum_{j=1}^{n} a_j^2.
\end{align}
When $n$-sequences are represented as points in an $n$-dimensional ($n$-D) space, the set
\begin{align}
\calA^\bullet = \left\{ a_1, a_2, \cdots, a_n \bigg| \sum_{j=1}^{n} a_j^2 \leq E^\bullet \right\}, \label{eq:sphereset}
\end{align}
consists of all amplitude sequences located in or on the surface of the $n$-sphere of squared radius $E^\bullet$.
The zero-energy point is at the center of this sphere.

The composition of a sequence $x^n \in \calA^n$ is defined as $\Comp = [n_1, n_2,\cdots,n_{\namp}]$, where $n_j$ denotes the number of times the $j^{\text{th}}$ element of $\calA$ occurs in $x^n$, i.e.,
\begin{align}
    n_j = \sum_{l=1}^n \mathds{1}[x_l=a_j],
\end{align}
for $j = 1, 2,\cdots, \namp$.
Here $\mathds{1}[\cdot]$ is the indicator function which is 1 when its argument is true, and 0 otherwise.
The number of unique $n$-sequences with the same composition $C$ is given by the multinomial coefficient 
\begin{align}\label{eq:multinom}
\text{MC}(\Comp) = \frac{n!}{\prod_{j=1}^{\namp} n_j!}.
\end{align}

For a set $\calAspd$ of amplitude sequences with $P_A(a)$ induced on $\calA$,
the average energy per symbol is
\begin{align} 
E  = \sum_{a\in\calA} P_{A}(a) a^2. \label{eq:aveng}
\end{align}
The shaping rate of the set $\calAspd$ is defined as
\begin{align}
    \Rs = \frac{\log_2  \left| \calAspd \right| }{n}, \label{eq:shapingrate}
\end{align}
in bit/1-D.
The input blocklength of a shaping algorithm that indexes sequences from the shaping set $\calAspd$ is
\begin{align}
k = \lf \log_2 \left( \left| \calAspd \right| \right) \rf, \label{eq:inputlength}
\end{align}
in bits.
It can be shown that the parameters of a shaping code $\calAspd$ satisfy the following inequality
\begin{align}
    \ent(A) \stackrel{(a)}{\geq} \frac{\log \left|\calAspd\right|}{n} \stackrel{(b)}{\geq} \frac{k}{n}, \label{eq:rloss_relations}
\end{align}
where (a) is due to the finite blocklength $n$ and (b) is due to the binary-input nature of the shaping algorithm, i.e., the rounding in \eqref{eq:inputlength}.
Here $\ent(A)$ is the entropy of $P_A$ in bits.
In \eqref{eq:rloss_relations}, both (a) and (b) are satisfied with equality when $n\rightarrow\infty$.
The rate loss of a shaping set $\calAspd$ with induced distribution $P_A(a)$ can then be defined in bit/1-D as
\begin{align}
\rloss = \ent\left(P_A\right) - \frac{k}{n}. \label{eq:rateloss}
\end{align}

\subsection{Shaping Architecture vs. Shaping Algorithm}\label{ssec:dm_architecture_vs_algorithm}
The aforementioned shaping schemes have in common that they are aiming at solving an indexing problem, which is that the binary input at the mapper determines an output sequence. At the receiver side, the inverse operation is carried out. This indexing problem has many different approaches to, and for proper characterization and categorization, it is insightful to differentiate between architecture and algorithm.  

When we speak of the architecture, we mean the underlying principle behind the mapping operation, which in turn can be realized with various different algorithms as shown in the fourth layer of Fig.~\ref{fig:shapingtaxonomy}. 
For instance, the CCDM principle (i.e., architecture) is that the sequences at the mapper output have a fixed number of occurrences of each amplitude, i.e., they satisfy the composition $\Comp$. 
Furthermore, the mapping algorithm can operate on one nonbinary or several binary subsets of the output sequence. Bit-level~\cite{SteinerSB2018_PDM,pikus2017} and parallel-amplitude~\cite{Fehenberger2019Arxiv_PASR} designs are modifications to the conventional CCDM architecture that carry out such a transformation from one nonbinary to several binary DMs. Among all algorithms, a lookup table (LUT) is probably the simplest way to solve the CCDM indexing problem, yet the LUT size table is prohibitively large as it reaches Gbit size already for short blocklengths \cite{Millar2019ECOC_HCSS}. The original mapping method for a nonbinary-alphabet CCDM is AC~\cite[Sec.~IV]{ccdm} which is modified from~\cite{ramabadran1990}. 
For binary-output CCDM, SR has recently been proposed as a low-serialism alternative to CCDM. 
MPDM~\cite{Fehenberger2019TCOM_MPDM} extends the CCDM principle (and thus architecture) by using variable-composition DM, yet internally uses CCDM methods for mapping and demapping. 

As another example, the \spsh~principle (i.e., architecture) is that the sequences at the output of the shaper satisfy a maximum-energy constraint, i.e., they satisfy \eqref{eq:sphereset}.
The problem of indexing these sequences can be solved again by a using a LUT.
On the other hand, ESS~\cite{willems1993}, SM~\cite{laroia1994} and~\cite[Algorithm 1]{laroia1994} are constructive algorithms to index sequences in a sphere.
The required storage and computational complexity of these algorithms are compared in Sec.~\ref{sec:complexity}.
For further discussion on SM, we refer the reader to~\cite{laroia1994,Khandani1993_ShapingMultiD_part1},~\cite[Ch. 8]{Tretter2002_ShapePrecodeITU} and~\cite[Sec. 4.3]{Fischer2002_PrecodingShaping}.

In this work, the architectures of CCDM and MPDM, and different algorithmic realizations of DM are discussed in Sec.~\ref{ssec:distribution_matchers}.
The architecture of \spsh~and different ways of realizing it (ESS and SM) are examined in Sec.~\ref{ssec:rate_matchers}.

\section{Signaling Schemes}\label{sec:signalingschemes}
\subsection{Uniform Signaling} 
In uniform signaling, a $k$-bit uniform sequence $u^k = (u_1,u_2, \cdots, u_k)$ is encoded by a rate $R_c = k/n_c$ FEC code, as shown in Fig.~\ref{fig:PASblockdiag} (a).
Afterwards, the coded sequence $c^{n_c}$ is divided into $m$-bit vectors, each of which is mapped to a channel input symbol via the symbol mapper.
Finally, assuming that $n_c/m=n$, the sequence $x^n \in \calS^n \times \calA^n$ is transmitted over the channel. 
The transmission rate of this construction is $R=k/n$~bit/1-D.
We will compare the uniform and shaped signaling techniques at the same transmission rate $R$, as it is obviously the only fair comparison as recently discussed in~\cite[Sec. IV-A]{Amari2019_IntroducingESSoptics} and~\cite{VassilievaKI2019_FairnessPASQAM}.

\subsection{Probabilistic Amplitude Shaping} \label{ssec:pas}
\begin{figure*}[t]
	\centering		
\includegraphics[width=1.1\columnwidth]{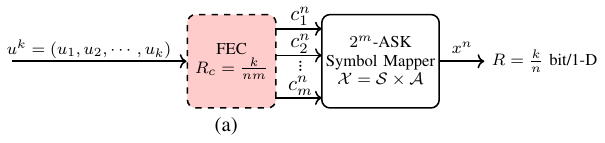}
\includegraphics[width=1.45\columnwidth]{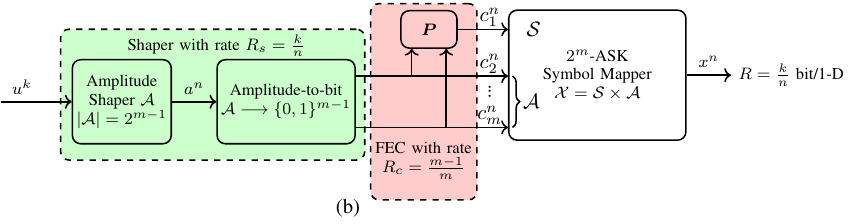}
\includegraphics[width=1.9\columnwidth]{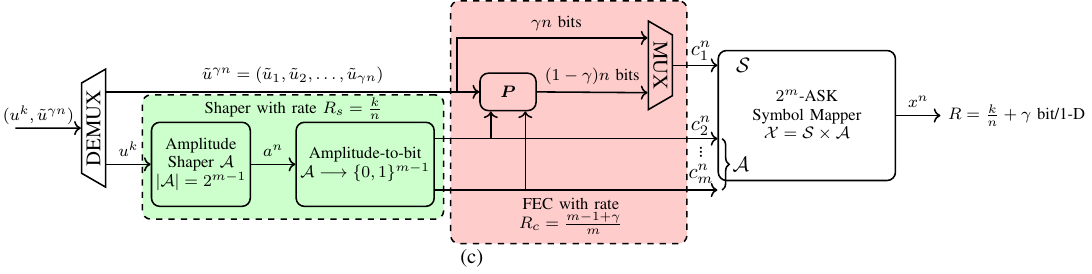}
	\caption{Signaling options: (a) uniform signaling, (b) PAS (all information is on amplitudes), (c) modified PAS (extra data is carried on signs).}
    \label{fig:PASblockdiag}
\end{figure*}

B\"{o}cherer {\it et al.} introduced in~\cite{bocherer2015} the PAS framework which couples an outer shaping code and an inner FEC code to realize {\it shaped and coded modulation}.
Figure~\ref{fig:PASblockdiag} (b) shows the basic PAS architecture where first, an amplitude shaping block maps a $k$-bit uniform information sequence $u^k$ to an $n$-amplitude sequence $a^n = (a_1, a_2,\cdots, a_n)$ in an invertible manner, where $a_j \in \calA$ for $j = 1,2,\cdots,n$.
After this mapping block, these amplitudes are transformed into bits using the last $m-1$~bits of the employed binary labeling.
We note that due to the shaped nature of $a^n$, the bits at the output of the amplitude-to-bit conversion in Fig.~\ref{fig:PASblockdiag} (b) are nonuniform.
These $n(m-1)$ nonuniform bits $c^n_2, c^n_3, \cdots, c^n_m$ are then used as the input of a systematic, rate $R_c=(m-1)/m$ FEC code which is specified by an $n(m-1)$-by-$nm$ parity-check matrix $\boldP$.
The $n$-bit parity output of this code is employed as the sign bit-level, i.e., the first bit of the binary labels, to determine the sign sequence $s^n = (s_1, s_2, \cdots, s_n)$.
Finally, $x^n = s^n a^n \in \calS^n \times \calA^n$ is transmitted over the channel.
The transmission rate of this scheme is $R = k/n$~bit/1-D.

To use a higher FEC code rate\footnote{Since symbol-level shaping strategies determine $m-1$ amplitude bits prior to FEC encoding, they can only be combined with FEC code rates $R_c \geq (m-1)/m$. To employ lower FEC code rates $R_c < (m-1)/m$, bit-level shaping strategies which only determine a subset of $m-1$ amplitude bits should be employed as in~\cite{pikus2017,SteinerSB2018_PDM,Gultekin2019Arxiv_PESS}.} $R_c > (m-1)/m$, a modified PAS architecture is proposed in~\cite{bocherer2015} as shown in Fig.~\ref{fig:PASblockdiag} (c).
The code rate in this sceme is $R_c = (m-1+\gamma)/m$ where $\gamma = R_cm-(m-1)$ sepecifies the number of extra data bits that will be transmitted per symbol.
In this modified structure, in addition to the $n(m-1)$~bit output of the shaper, extra $\gamma n$ information bits $\tilde{u}^{\gamma n}$ are fed to the FEC code which is now specified by an $(m-1+\gamma)n$-by-$mn$ parity-check matrix $\boldP$.
The $(1-\gamma)n$~bit parity output of the FEC code is then multiplexed with the uniform bits $\tilde{u}^{\gamma n}$ to form an $n$-bit sequence that will select the signs.
The transmission rate of this scheme is $R = k/n + \gamma$~bit/1-D.

\begin{example}[{\bf Shaping, FEC and transmission rates in PAS}]\label{ex:pasvalues}
Consider the PAS architecture with 8-ASK, a rate $R_c=5/6$ FEC code, and a target rate $R=2.25$~bit/1-D.
The rate of the extra data that will be carried in the signs of the channel inputs is $\gamma = R_cm-(m-1) = 0.5$~bit/1-D.
Therefore the rate of the amplitude shaper should be $k/n = R - \gamma = 1.75$~bit/1-D.
If the length of the FEC code is $n_c=648$~bits, the blocklength is $n=n_c/m=216$~real symbols.
Then the output set of the amplitude shaper must consist of at least $2^{k}=2^{216\cdot1.75}=2^{378}$ sequences.
\end{example}

\subsection{PAS Receiver}\label{ssec:pasrec}
At the receiver, the log-likelihood ratio (LLR) of $B_j$ is computed by a soft demapper as
\begin{eqnarray}
L(B_j) = \log \left( \frac{\sum_{x\in \calX_{j,0}} P_X(x) f_{Y|X}(y|x)}{\sum_{x\in \calX_{j,1}} P_X(x) f_{Y|X}(y|x)} \right) \label{eq:llr}
\end{eqnarray}
based on the channel output $Y$ for $j = 1, 2,\cdots,m$, where $\calX_{j,u}$ denotes the set of $X\in\calX$ which have $B_j=u$ in their binary labels for $u  \in \{0, 1\}$.
We emphasize that the nonuniform a-priori information on the symbols is used in \eqref{eq:llr}.
Instead of symbol-wise probabilities $P_X(x)$, bit-wise probabilities $P_{B_j}(b_j)$ for $j = 1, 2,\cdots, m$ can equivalently be used to compute the LLRs as in~\cite[eq.~(60)]{bocherer2015} or~\cite[eq. (3.29-32)]{Szczecinski2015_BICMbook}.
Then based on the LLRs, a binary FEC decoder recovers the bits that were encoded by the FEC code.
In the case of uniform signaling, these bits are the estimates of the information bits.
For the PAS architecture shown in Fig.~\ref{fig:PASblockdiag} (b), the output of the decoder consists of the estimates of the amplitude bits. 
Then these are mapped back to the information bit estimates using the inverse functions of the blocks in the shaper (green box), i.e., the corresponding bit-to-amplitude mapper followed by the corresponding amplitude deshaper.
In addition to this, for the PAS architecture shown in Fig.~\ref{fig:PASblockdiag} (c), the decoder also outputs the estimates of the $\gamma n$ extra data bits which were used as some of the signs.
According to~\cite{bocherer2015}, a bit-metric decoder achieves the rate {$\rbmd$} for any input distribution $P_X(x)$,
\begin{eqnarray}
\rbmd &=& \left[ \ent(X) - \sum_{j=1}^{m} \ent(B_j|Y ) \right]^+, \label{eq:rbmd}
\end{eqnarray}
where $[\cdot]^+ = \max\{0,\cdot\}$.

\subsection{Selection of Parameters for PAS} \label{ssec:PASparameterselection}
In this section, we study the optimum shaping and FEC coding rates for PAS using AIRs.
Thus we consider the case where $n\rightarrow\infty$ which implies that $k=n\ent(A)$ from~\eqref{eq:rloss_relations}, and consequently, $R = \ent(A) +\gamma$.

\begin{figure}[t]
\centering		
\includegraphics[width=\columnwidth]{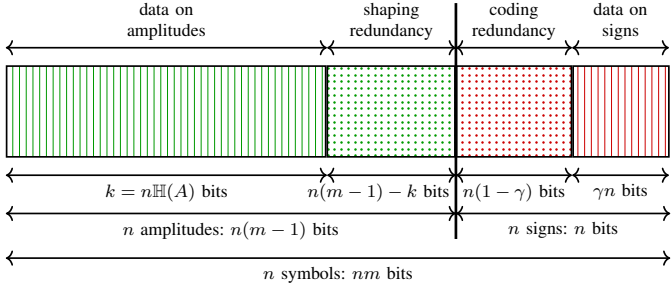}
\caption{Content of a channel input sequence produced by PAS.}
\label{fig:PASsequence}
\end{figure}

In the PAS architecture, to obtain a target rate $R = \ent(A) + \gamma$ using the $2^m$-ASK constellation, a total of $n(m-R)$ redundancy bits are added to a channel input sequence by shaping and coding operations combined.
Shaping is responsible for $n(m-1-\ent(A))$ redundant bits whereas coding adds $n(\ent(A)+1-R)$.
This is illustrated in Fig.~\ref{fig:PASsequence} where the content of a channel input sequence produced by the generalized PAS architecture of Fig.~\ref{fig:PASblockdiag} (c) is shown.
The striped areas represent the information carried in signs (red) which is $\gamma n$~bits, and in amplitudes (green) which is $k=n\ent(A)$~bits.
Dotted areas show the redundant bits in a sequence.
When $\gamma=0$, i.e., $R_c = (m-1)/m$, all signs are selected by redundancy bits and thus, the striped red area in Fig.~\ref{fig:PASsequence} vanishes.
When $\ent(A) = m-1$, the amplitudes are uniformly distributed, i.e., there is no shaping, and thus, the dotted green area in Fig.~\ref{fig:PASsequence} disappears.
We note that a similar illustration was provided for a single ASK symbol in~\cite[Fig. 9]{Fehenberger2019TCOM_MPDM}.
In Table~\ref{tab:pasexamplevalues}, the content of a sequence at the output of a PAS transmitter (in accordance with Fig.~\ref{fig:PASsequence}) is tabulated for Example~\ref{ex:pasvalues} where $n=216$.

\begin{table}[ht]
\renewcommand{\arraystretch}{1.3}
\centering
\caption{Content of an amplitude sequence as in Fig.~\ref{fig:PASsequence} based on Example~\ref{ex:pasvalues}}
\resizebox{\columnwidth}{!}{
\begin{tabular}{c|c|c|c} 
Parameter & \begin{tabular}{@{}c@{}} Formula \\ (per $n$-sequence)\end{tabular} & \begin{tabular}{@{}c@{}} Value per 1-D \\ (Example~\ref{ex:pasvalues}) \end{tabular} & \begin{tabular}{@{}c@{}} Value per 216-D \\ (Example~\ref{ex:pasvalues}) \end{tabular}  \\
\hline\hline
Data on amp. & $n\ent(A)$ & 1.75 & 378 \\
Data on sign & $n\gamma$ & 0.50 & 108 \\
\hline
Shap. redundancy & $n(m-1-\ent(A))$ & 0.25 & 54 \\
Cod. redundancy & $n(\ent(A)+1-R)$ & 0.50 & 108 \\
\hline
Redundancy & $n(m-R)$ & 0.75 & 162 \\
Data, $nR$ & $n(\ent(A)+\gamma)$ & 2.25 & 486\\
\end{tabular}
}
\label{tab:pasexamplevalues}
\end{table}

When the input is constrained to be MB-distributed, $\ent(X)=\ent(A)+1$ can be used as a design parameter which tunes the balance between shaping and coding redundancies at a fixed rate $R$.
More specifically, the entropy $\ent(A)$ of the MB distribution is controlled by $\lambda$. 
Thus by changing $\lambda$, the amount of shaping redundancy in an amplitude can be adjusted. The question is then how to choose the optimum $\lambda$.
Following Wachsmann, Fischer and Huber~\cite{FischerHW1996_MultilevelCoding,wachsmann1999}, we use the gap to capacity (normalized SNR), which is defined as
\begin{eqnarray}
\delsnr = \frac{ \text{required SNR such that $\rbmd = R$} }{2^{2R}-1} , \label{gapeq}
\end{eqnarray}
as the metric to be minimized when searching for the optimum MB distribution\footnote{In general, the gap-to-capacity curve can be plotted for any parametric family of distributions. Here we only consider the MB distributions since they have been shown to perform very close to the capacity of ASK constellations over the AWGN channel and maximize the energy efficiency~\cite{kschischang1993}.} for a fixed rate $R$ and constellation size $2^m$.
The numerator in \eqref{gapeq} is the SNR value at which $\rbmd=R$ for a given $P_X$, and the denominator is the SNR value at which the capacity $C=R$.
We note that instead of the MI in~\cite[eq. (55)]{wachsmann1999}, we now use the BMD rate of~\eqref{eq:rbmd}.
Observing from Fig.~\ref{fig:PASsequence} that $1-\gamma=1-(R-\ent(A))$, the rate of the FEC code that should be employed in PAS to obtain a transmission rate $R$ for a given constellation entropy $\ent(X)$ is
\begin{align}
R_c = \frac{m-1+\gamma}{m} = \frac{m+R-\ent(A)-1}{m} = \frac{m+R-\ent(X)}{m}. \label{eq:coderate}
\end{align}

\begin{figure}[t]
\centering
\includegraphics[width=\columnwidth]{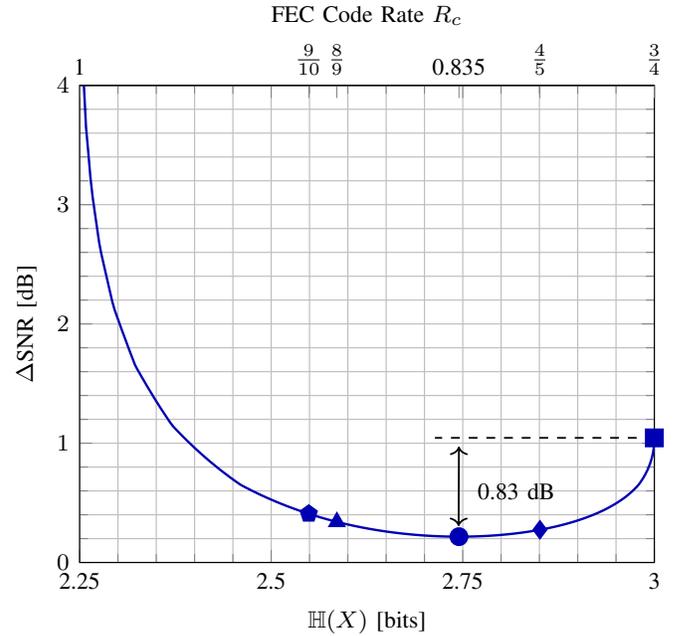}
\caption{Channel input entropy vs. gap-to-capacity for 8-ASK at the target rate of $R=2.25$~bit/1-D. 
The x-axis above shows the corresponding FEC code rates.}
\label{fig:G2C}
\end{figure}

\begin{example}[\bf Optimal PAS parameters]\label{ex:optimalPAS}
In Fig.~\ref{fig:G2C}, the entropy $\ent(X)$ of an MB-distributed input $X$ with $|\calX|=8$ two-sided amplitude levels (i.e., 8-ASK) vs. $\delsnr$ is plotted for $R=2.25$~bit/1-D. 
On the top horizontal axis, the corresponding FEC code rates in \eqref{eq:coderate} are also shown.
The rightmost point (indicated by a square) corresponds to uniform signaling where the target rate of 2.25~bit/1-D is obtained by using a FEC code of rate $R_c=R/m=3/4$. In this trivial case, all 0.75~bits of redundancy are added by the coding operation, and the gap to capacity $\delsnr$ is 1.04 dB.
The leftmost part of the curve where $\ent(X)$ goes to $R$ belongs to the uncoded signaling case, i.e., $R_c = 1$, where $R$ is attained by shaping the constellation such that $\ent(X)=R$.
Here $\delsnr$ is infinite since without coding, reliable communication is only possible over a noiseless channel.
The minimum $\delsnr$ in Fig.~\ref{fig:G2C} is obtained with $\ent(X)=2.745$, which corresponds to $R_c=0.835$ from \eqref{eq:coderate}. 
In IEEE DVB-S2~\cite{DVBS2} and 802.11~\cite{80211-2016}, the code rate that is closest to 0.835 is $5/6\approx0.833$. 
Accordingly, the best performance is expected to be provided by FEC rate $5/6$, with an SNR gain over uniform that amounts according to this analysis to 0.83~dB. This will be confirmed by the numerical simulations presented in Sec.~\ref{end2end}.
\end{example}

\section{Distribution Matching and Sphere Shaping Schemes}\label{sec:shaping_schemes}
This section gives an overview of various shaping schemes that are compatible with the PAS framework. We focus on constructive methods, i.e., the direct use of a LUT for mapping or demapping is not considered herein due to its impracticality even for moderate blocklengths. 
Also, only fixed-length schemes are considered.
      
\subsection{Distribution Matching Schemes (Direct Method)}\label{ssec:distribution_matchers}
In the following, an overview of distribution matching architectures and algorithms is given. The difference between these two aspects was discussed in Sec.~\ref{ssec:dm_architecture_vs_algorithm}. All of the following schemes have in common that a certain PMF is targeted explicitly. For finite-length DM, this means that some quantization might be required as to achieve an integer-valued composition. Possible quantization rules include a simple rounding operation~\cite[Sec. V-A2]{bocherer2015}, or minimizing the Kullback-Leibler divergence \cite{Bocherer2016TransIT_OptimumQuantize}. We note that neither of these approaches is necessarily optimal in achieving the maximum information rate for a given $n$ and channel law. 

CCDM has been proposed for PAS in~\cite{ccdm}.
We speak of constant composition if all matcher output sequences are permutations of a particular base sequence, which is typically described by the composition $\Comp$ stating the number of occurrences of each amplitude.
    The number of unique output sequences of the corresponding matcher, i.e., the cardinality of the shaping set $\calA^\circ \subseteq \calA^n$, is given by the multinomial coefficient \MultinomCoeff{\Comp}, as defined in \eqref{eq:multinom}.
    Each amplitude sequence in $\calA^\circ$ has the same energy $E^\circ$, and consequently, they all are located on the $n$-shell of squared radius $E^\circ$ as shown in Fig.~\ref{fig:shell2sphere}.
    
    \begin{example}[{\bf CCDM}]\label{ex:ccdm}
    We consider the target PMF $P_A = [0.4378, 0.3212, 0.1728, 0.0682]$ over $\calA = \{1, 3, 5, 7\}$ with $\ent(A)=1.75$. 
    The composition that is obtained for $n=216$ with the quantization rule proposed in~\cite[Algo. 2]{Bocherer2016TransIT_OptimumQuantize} is $C = [95, 69, 37, 15]$.
    The shaping rate \eqref{eq:shapingrate} of the matcher that produces sequences with composition $C$ is $R_s=1.6991$~bit/1-D.
    The input length \eqref{eq:inputlength} of this matcher is $k=367$~bits.
    \end{example}
 
      MPDM has been proposed in \cite{Fehenberger2019TCOM_MPDM} as an extension to CCDM that lifts the constant-composition principle. 
      MPDM is based on the idea that the target distribution ${\Comp}$ need not be achieved in each output sequence; rather, it is sufficient if the ensemble average over all sequences gives the target composition. 
      Considering the example of pairwise partition in~\cite{Fehenberger2019TCOM_MPDM}, this means that each composition has a complement, both with the same number of occurrences, such that their average is the target distribution. 
      There are, however, no known constructive algorithm for this variable-composition mapping problem. 
      This is circumvented by reducing the number of unique sequences of each composition to be a power of two, which can come at the expense of some small rate loss. 
      This additional constraint enables Huffman coding on the compositions, i.e., we can build a tree where a variable-length prefix determines the node and thus, the composition to be used. 
      The remaining binary payload is mapped with conventional CCDM techniques. 
      Note that the prefix and payload length are balanced such that the overall mapping operation is fixed-length. 
      It has been shown in~\cite{Fehenberger2019TCOM_MPDM} that pairwise MPDM with such a tree structure gives an approximately fourfold length reduction compared to CCDM at the same information rate. 
      It has also been demonstrated to give significant rate improvements for a fixed block fixed length for various QAM formats transmitted over the AWGN channel \cite{Millar2019JLT_MPDM} and the optical fiber channel \cite{Fehenberger2019OFC_MPDM}.
      
      \begin{example}[{\bf MPDM}]
      We consider the same target PMF as in Example~\ref{ex:ccdm}.
      Pairwise MPDM with tree structure utilizes 945 compositions whose average is again $[95, 69, 37, 15]$.
      The shaping rate \eqref{eq:shapingrate} of the matcher that produces sequences with these compositions is $R_s=1.7315$~bit/1-D.
      The corresponding input length \eqref{eq:inputlength} is $k=374$ which is 7 bits more than that of CCDM which is a 1.9\% rate increase.
      \end{example}

    CCDM has initially been realized with AC, which is sequential in the input length, i.e., at most $k$ serial operations have be carried out for mapping and $n$ for demapping\footnote{Note that this describes the worst-case serialism if the DM operation cannot be terminated early, which could be the case when the remainder of the output sequence follows at some point with certainty. Also, this metric does not incorporate the complexity of the computations inside each step as discussed in Sec.~\ref{sec:complexity}.}. Since the serialism of the AC method can be challenging to achieve for high-throughput CCDM operation, means to run several DMs in parallel have been proposed. 
    For BL-DM~\cite{pikus2017} or PDM~\cite{SteinerSB2018_PDM} where the target distribution is a product distribution, the parallelization factor is $\log_2\namp=m-1$ since one binary-alphabet DM is used for each bit level. This approach has been numerically shown to have reduced rate loss compared to a single nonbinary DM, yet comes at the expense of having the DM output limited to compositions that are generated from a product distribution. 
    In \cite{Fehenberger2019Arxiv_PASR}, a different parallelization technique has been proposed, which operates on amplitudes rather than on bit levels. 
    For each of the $\namp-1$ out of $\namp$ amplitudes, a binary-alphabet DM is operated in parallel, with the first DM determining the position of the first amplitude, the second DM where to position the second amplitude within those positions that have not been occupied by the preceding (i.e., first) amplitude. These DM operations can be run in parallel and only the final step of combining the subsequences into the nonbinary output sequence is sequential. We note that both bit-level DM and parallel-amplitude DM are compatible with MPDM.
    
    The schemes discussed in the preceding paragraphs can be considered as extensions to the CCDM architecture that either nest various CCDMs for improved performance (MPDM) or transform a nonbinary CCDM into several binary CCDMs to achieve a larger parallelization (bit-level and parallel-amplitude DM). 
    In \cite{Fehenberger2019Arxiv_PASR}, SR has been proposed as an alternative to the conventional AC algorithm for CCDM as shown in the bottom layer of Fig.~\ref{fig:shapingtaxonomy}. 
    SR solves the CCDM indexing problem by representing a binary-alphabet sequence as a constant-order subset that determines the position of either binary symbol. 
    For a given sorting, such as lexicographical, the rank of such a subset is found by ``enumerating" all preceding sequences which is used for source coding in~\cite{schalkwijk1972,cover1973} and for shaping in~\cite{willems1993,GultekinWillems2018ISIT_EnumerativeShaping}. 
    This mapping from sequence to (binary) rank is called unranking in the combinatorics literature and acts as inverse mapping. 
    The ranking operation from bits to sequence is DM mapping. 
    The advantage of SR over AC is that the number of serial operations is significantly reduced~\cite[Sec.~V]{Fehenberger2019Arxiv_PASR}.
    
\subsection{Sphere Shaping Schemes (Indirect Method)}\label{ssec:rate_matchers}
In this section, a review of \spsh~algorithms is provided.
All ensuing algorithms target a certain rate, i.e., the number of unique output sequences, rather than a PMF.
To this end, for a given $\calA$, $n$ and target $k$, the maximum-energy constraint $E^\bullet$ is selected such that the set $\calA^\bullet$, as defined in \eqref{eq:sphereset}, satisfies $|\calA^\bullet| \geq 2^k$.
This set consists of all $2^m$-ASK amplitude lattice\footnote{We use ``$2^m$-ASK amplitude lattice" for the $n$-fold Cartesian product of $\{1, 3,\cdots, 2^m-1\}$ with itself, i.e., $\calA^n$.} points on the surface or in the $n$-sphere of square radius $E^\bullet$ as shown in Fig.~\ref{fig:shell2sphere}.
We note that possible sequence energy values for these points, i.e., squared radii of the $n$-dimensional shells that the sequences are located on, are $\left\{ n, n+8,\cdots, E^\bullet \right\}$, and the number of shells is calculated as ~\cite{Gultekin2019Arxiv_ESS}
\begin{eqnarray}
L = \lf \frac{E^\bullet-N}{8} \rf +1, \label{eq:L}
\end{eqnarray}.

\begin{remark}\label{rem:L_func_N}
We see from the sphere-hardening result discussed, e.g., by Wozencraft and Jacobs in~\cite[Sec. 5.5]{wozencraft1965}, that $E^\bullet \approx nE$ for large $n$. 
Following Laroia {\it et al.}~\cite[Sec. III-A]{laroia1994} and approximating the required average energy to transmit $R$ bit/1-D by $c2^{2R}$, we can write $E^\bullet \approx nc2^{2R}$ where $c$ is some constant. 
Therefore $L$ in \eqref{eq:L} is a linear function of $n$ for a fixed rate $R$.
\end{remark}

\begin{example}[{\bf Sphere shaping}]\label{ex:sphereshaping}
The shaping set $\calA^\bullet \subset \calA^n$ for the parameters $n=64$, $\calA = \{ 1, 3, 5, 7 \}$ and $E^\bullet=768$, i.e., $L=89$, has the shaping rate $\Rs=1.7538$~bit/1-D.
The input length of the corresponding amplitude shaper is $k=112$~bits.
The induced PMF is $P_A(a) = [0.42, 0.32, 0.18, 0.08]$ over $\calA$, where the average energy per dimension is $E=11.6316$.
\end{example}

In the following, we explain two different algorithms to realize \spsh: Enumerative sphere shaping (ESS) and shell mapping (SM).
Provided with identical parameters, these two address the same set $\calA^\bullet$ of sequences where the difference is in the bits-to-amplitudes mapping.

ESS starts from the assumption that the energy-bounded amplitude sequences, i.e., $a^n\in\calA^\bullet$, can be ordered lexicographically.
Thus the index of an amplitude sequence is defined to be the number of sequences which are lexicographically smaller.
To represent $n$-amplitude sequences in a sphere, an energy-bounded enumerative amplitude trellis is constructed~\cite[Sec.~III-B]{Gultekin2019Arxiv_ESS}.
Operating on this enumerative trellis, $n$-step recursive algorithms are devised to realize the lexicographical index-sequence mapping in an efficient manner~\cite{willems1993,GultekinWillems2018ISIT_EnumerativeShaping}.
These algorithms demand the storage of a matrix (i.e., the trellis) of size $(n+1)\times L$ where each element can be up to $\lc nR_s \rc$-bit long.
The required storage and computational complexity of ESS is discussed in Sec.~\ref{sec:complexity}.

Another way of ordering $n$-amplitude sequences in a sphere is to sort them based on their energy, i.e., based on the index of the $n$-dimensional shell that they are located on.
Sequences on the same shell can be sorted lexicographically.
To this end, a trellis which is different from that of ESS is constructed~\cite{laroia1994,GultekinWillems2018ISIT_EnumerativeShaping}.
Based on this trellis, two different indexing algorithms are proposed in~\cite{laroia1994}.
The first one~\cite[Algorithm 1]{laroia1994}, which was proposed around the same time as ESS~\cite{willems1993}, has performance and complexity similar to ESS. 
The second one~\cite[Algorithm 2]{laroia1994}, which is the well-known shell mapping (SM), is based on the divide-and-conquer (D\&C) principle, and enables a tradeoff between the computational and storage complexities~\cite[Sec. 4.3]{Fischer2002_PrecodingShaping}.
The D\&C principle was used to enumerate sequences from the Leech lattice earlier in~\cite{LangL1989_LeechLatticeModem}.
The basic principle is to successively divide an $n$-dimensional indexing problem into two $n/2$-dimensional problems, creating a $\log_2n$-step operation.
Consequently, SM demands the storage of a matrix of size $(\log_2n+1)\times L$ where each element is again $\lc nR_s \rc$-bit long.
The required storage and computational complexity of SM is discussed in Sec.~\ref{sec:complexity}.
      
In their initial proposals, shaping matrices of ESS~\cite{willems1993}, SM~\cite{laroia1994} and~\cite[Algorithm 1]{laroia1994} are computed with full-precision (FP).
To decrease the storage complexity of ESS and SM, a bounded-precision (BP) implementation method is proposed in~\cite{GultekinWillems2018ISIT_EnumerativeShaping}.
The idea is that any number can be expressed in base-2 as $m\cdot 2^p$.
Here $m$ and $p$ are called the mantissa and the exponent, stored using $n_m$ and $n_p$~bits, respectively.
Then each number in a shaping matrix, i.e., in the trellis, is rounded down to $n_m$~bits after its computation, and stored in the form $(m,p)$.
The invertibility of ESS and SM functions is preserved with this approach~\cite{GultekinWillems2018ISIT_EnumerativeShaping}.
We note that the BP implementation can also be used to realize~\cite[Algorithm 1]{laroia1994}.
The BP implementation decreases the memory required to store an element of the shaping trellis from $\lc nR_s \rc$~bits to $n_m+n_p$~bits. 
Typical values of $n_m$ and $n_p$ are a few bytes.
The required storage and the computational complexity of BP implementation is discussed in Sec.~\ref{sec:complexity}.
The disadvantage of this approximation is that the numbers in the trellis, and thus the number of output sequences decreases, causing a rate loss.
However this rate loss is shown to be upper-bounded by $-\log_2(1-2^{1-n_m})$ bit/1-D~\cite{GultekinWillems2018ISIT_EnumerativeShaping}.
    
\begin{example}[{\bf Bounded-preceision rate loss}]
If the shaping set $\calA^\bullet$ in Example~\ref{ex:sphereshaping} is constructed with BP using $n_m=9$~bit mantissas and $n_p=7$~bit exponents, the resulting rate loss is upper-bounded by 0.0056 bit/1-D.
For ESS and SM, the actual rate losses are 0.0021 and 0.0003 bit/1-D, respectively.
Since the shaping rate with FP was $R_s=1.7538$, these rate losses keep $R_s>1.75$, and consequently, keep $k=112$.
Therefore, we claim that when more than a few bytes are used to store mantissas, BP rate loss is smaller than the loss due to the rounding operation in \eqref{eq:inputlength}. 
Consequently, the operational rate $k/n$ is not affected.
However, the required memory to store an element of the shaping matrix drops from $\lc n\Rs \rc=113$ bits to $n_m+n_p=16$.
\end{example}

Both ESS and SM index the same set of sequences for fixed $n$, $\calA$ and $E^\bullet$.
The difference is in (i) the way algorithms are implemented  and (ii) the way the sequences are ordered.
We discuss the former difference in Sec.~\ref{sec:complexity}.
Due to the round-down operation in \eqref{eq:inputlength}, only the sequences with indices smaller than $2^k$ are actually utilized.
The remaining ones, i.e., the ones at the end of the list, are unused.
For SM, all these sequences have the highest possible energy $E^\bullet$.
On the other hand for ESS, these sequences are at the end of the lexicographical list and are not necessarily from the outermost shell.
Thus operationally, the output average symbol energy of SM is no greater than that of ESS, for a fixed set of parameters.
This difference could be important for ultra short blocklengths, however, for blocklengths larger than a few dozens, it becomes insignificant\footnote{The loss in power efficiency $P_{\text{loss,dB}}$ (in dB) can roughly be written as $P_{\text{loss,dB}} \approx 6\rloss$. We will later see in the context of Fig.~\ref{rlosses} (left) that the difference in rate losses of ESS and SM is only notable for blocklengths below a couple of dozens, and so does the loss in power efficiency.}.
Furthermore, as discussed in~\cite{GultekinW2019_OptimumTrellis}, by simply removing some connections from the shaping trellis, it is possible to force the discarded sequences to be from the outermost shell for ESS as well.

\subsection{Geometric Interpretation of the Shaping Approaches}\label{ssec:signalspace_comparison}
\begin{figure}[t]
\centering
\includegraphics[width=\columnwidth]{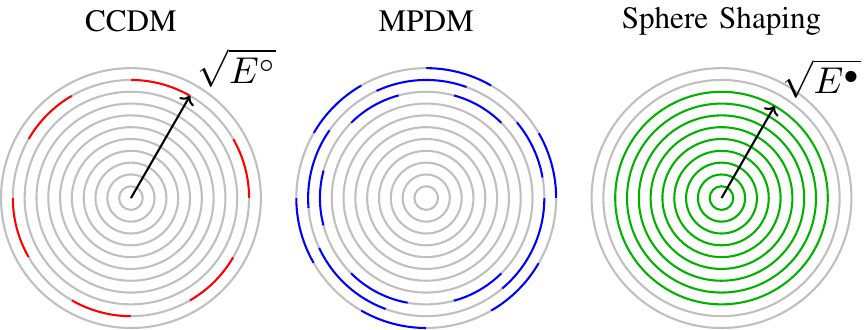}
\caption{The illustration of the employed $n$-dimensional signal points by CCDM (left), MPDM (middle) and \spsh~(right). Each circle represents an $n$-dimensional shell. Darker portions of the shells indicate the signal points on them which are utilized by the corresponding shaping approach.}
\label{fig:shell2sphere}
\end{figure} 

Output sequences of CCDM have a fixed composition and thus, all have the same sequence energy $nE$, i.e., they are located on the $n$-dimensional shell of squared radius $E^\circ=nE$.
We note that there are multiple compositions that lead to the same sequence energy and thus, the corresponding shell is only partially utilized by CCDM, as shown in Fig.~\ref{fig:shell2sphere}~(left).
With multiple compositions at its output, MPDM makes use of multiple partly filled $n$-shells, as in Fig.~\ref{fig:shell2sphere}~(middle).
The average symbol energy as well as the square radius $E^\circledcirc$ of the outermost shell that is utilized by MPDM depend on the actual set of considered compositions. 
Finally, $n$-sphere shaping employs all sequences inside the $n$-dimensional sphere of squared radius $E^\bullet$, as shown in Fig.~\ref{fig:shell2sphere}~(right). Note that for simplicity, we have in this explanation neglected the constraint that any practical binary scheme can only address a power-of-two number of shaped sequences.
When all three approaches enclose the same number of sequences at a fixed $n$, their average energy as in \eqref{eq:aveng} satisfy $E_{\text{ccdm}} \geq E_{\text{mpdm}} \geq E_{\text{spsh}}$.
Thus at any blocklength, \spsh~makes use of the set of sequences having the least average energy and is thus the most energy-efficient scheme.
This observation will later be confirmed by the rate loss analysis in Sec.~\ref{ssec:rloss}.

\section{Performance Comparison}\label{sec:performance_comparison}
This section studies the performance of the shaping schemes explained in Sec.~\ref{sec:shaping_schemes}. 
The used metrics are (i) finite-length rate loss at a fixed blocklength $n$, (ii) maximum AIR for BMD and (iii) FER.
 
\subsection{Rate Loss Analysis}\label{ssec:rloss}
\begin{figure}[t]
\centering		
\includegraphics[width=\columnwidth]{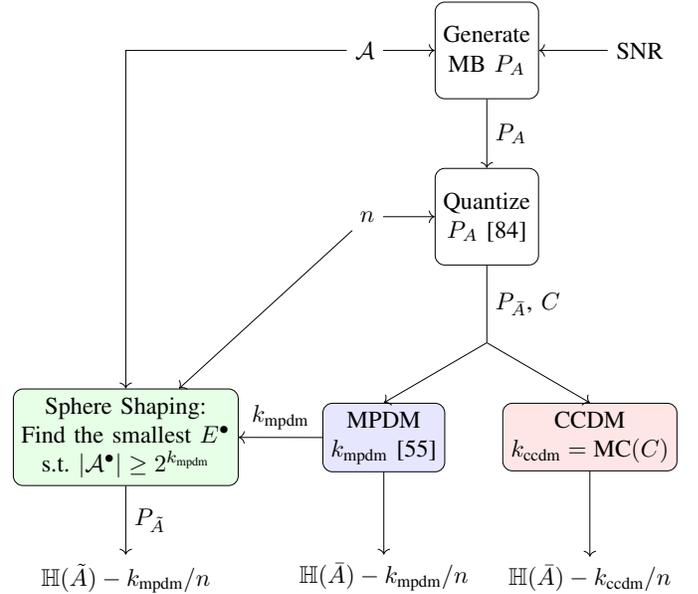}
\caption{Flowchart for the computation of rate loss for CCDM, MPDM and SPSH.}
\label{fig:rlossflowchart}
\end{figure}

The methodology of computing the rate loss for DM and \spsh~schemes in a fair manner is illustrated in Fig.~\ref{fig:rlossflowchart}.
For the DM schemes of Sec.~\ref{ssec:distribution_matchers}, the following steps are carried out in order to obtain the rate loss for a particular $n$. 
First, the target distribution $P_A$ (and thus the modulation order $2^m$) is fixed. 
The target distribution is MB, optimized for a particular SNR. 
We then quantize $P_A$ to $P_{\bar{A}}$ to get the integer-valued target composition $\Comp = nP_{\bar{A}}$, where the quantization criterion is to minimize the Kullback-Leibler divergence between $P_A$ and $P_{\bar{A}}$~\cite{Bocherer2016TransIT_OptimumQuantize}. 
For CCDM, $k=\floor{\log_2 \MultinomCoeff{C}}$ bits can be addressed where $\MultinomCoeff{\cdot}$ is as defined in \eqref{eq:multinom}. 
For nonconstant composition DMs such as MPDM \cite{Fehenberger2019TCOM_MPDM}, the number of addressable bits $k$ depends on the addressable bits of all constituent compositions, considering the specific constraints of the DM construction such as pairwise partitioning~\cite[Sec.~III-A]{Fehenberger2019TCOM_MPDM}.
The rate loss is finally computed as $\rloss=\ent(\bar{A})-k/\N$, as defined in \eqref{eq:rateloss}.

For the \spsh~schemes of Sec.~\ref{ssec:rate_matchers}, the approach must be different since it is not possible to explicitly target a certain distribution or composition.  
From the above methodology for MPDM schemes, we obtain the number of input bits $k$ for a given $n$. 
For each $n$, we find the smallest $E^\bullet$ (i.e., the squared radius of the sphere) such that the number of sequences inside the $n$-sphere $\calA^\bullet$ satisfies $\log_2(|\calA^\bullet|) \geq k$. 
We compute the induced distribution $P_{\tilde{A}}$~\cite[eq.~(17)]{Gultekin2019Arxiv_ESS}, and corresponding entropy $\ent(\tilde{A})$. 
The rate loss is again obtained as $\rloss=\ent(\tilde{A})-k/\N$. 
This procedure ensures that the DM and \spsh~schemes are compared at the same rate\footnote{For \spsh~schemes, the input length $k$ of CCDM can also be targeted during the rate loss computation. However, we prefer to use input length $k$ of MPDM since in general, $k_{\text{mpdm}} \geq k_{\text{ccdm}}$.}, i.e., at identical $k$ and $n$.
We note, however, that the target distribution and thus the source entropy differ.  

\begin{figure*}[t]
\centering
\includegraphics[width=\columnwidth]{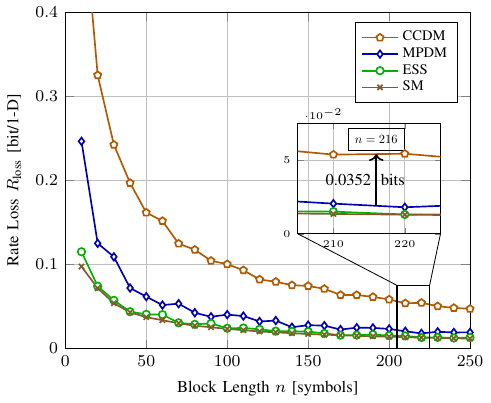}
\includegraphics[width=\columnwidth]{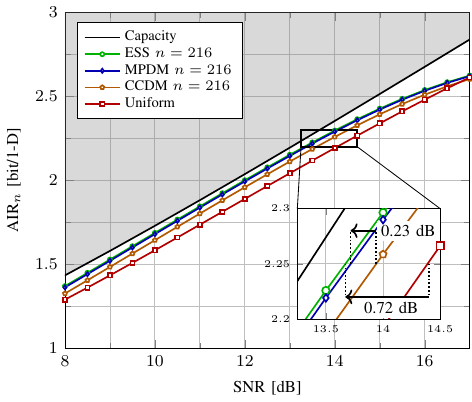}
\caption{(Left) Rate loss in \eqref{eq:rateloss} vs. blocklength $n$ for various shaping schemes. (Right) AIR in bit/1-D vs. SNR for BMD and 8-ASK. The rate losses for ESS and MPDM at $n=216$ are 0.0149 and 0.0215 bit/1-D, respectively.}
\label{rlosses}
\end{figure*}

\begin{example}[{\bf Rate loss comparison}]\label{ex:RLossMethod}
We consider the target distribution $P_A = [0.4378, 0.3212, 0.1728, 0.0682]$, with entropy $\ent(A) = 1.75$.
The $n$-type distribution that has the minimum informational divergence from $P_A$ for $n=216$ is $P_{\bar{A}} = [0.4398, 0.3194, 0.1713, 0.0694]$.
The corresponding composition is $C = [95, 69, 37, 15]$.
Starting with the same target distribution, i.e., with the same composition, the number compositions that are employed by MPDM is 945~\cite[Sec.~III-A]{Fehenberger2019TCOM_MPDM}. 
Since MPDM's set of compositions consists of pairs whose average is $C$, the induced distribution $P_{\bar{A}}$, its entropy $\ent(\bar{A})$ and the average symbol energy $E$ are the same as CCDM's.
The smallest $E^\bullet$ that gives $|\calA^\bullet| \geq 2^{k}$ is $E^\bullet=2376$ where $k$ is the input length of MPDM.
The corresponding induced distribution is $P_{\tilde{A}} = [0.4393, 0.3220, 0.1722, 0.0665]$.
Table~\ref{tab:rlossexample} shows the input length $k$, average symbol energy $E$ and rate loss $\rloss$ of CCDM, MPDM and ESS for these parameters.
We see that MPDM is able to address a larger set of sequences than CCDM, leading to a seven bit increase in the input length.
Since their induced distributions are the same, this is reflected as a decrease in rate loss.
Then starting with the same target $k$, ESS employs a set of sequences with smaller average energy.
This is also translated to a decrease in rate loss as shown in Table~\ref{tab:rlossexample}.

\begin{table}[ht]
\renewcommand{\arraystretch}{1.3}
\centering
\caption{Parameters Computed in Example~\ref{ex:RLossMethod}}
\begin{tabular}{cccccc}
Architecture & $k$ & $k/n$ & $E$ & $\ent(\bar{A})$ or $\ent(\tilde{A})$ & $\rloss$ \\
\hline
CCDM & 367 & 1.6991 & 11.00 & 1.7504 & 0.0513 \\
MPDM & 374 & 1.7315 & 11.00 & 1.7504 & 0.0189 \\
ESS  & 374 & 1.7315 & 10.90 & 1.7448 & 0.0133 \\
\end{tabular}
\label{tab:rlossexample}
\end{table}
\end{example}

Figure~\ref{rlosses} (left) shows rate loss vs. blocklength for CCDM, MPDM, ESS, and SM. 
The target distribution is the same as Example~\ref{ex:RLossMethod}. 
The target $k$ for ESS is the number of bits achieved by MPDM at each $n$ which ensures the same transmission rate $R$. 
We observe that all advanced schemes, i.e., MPDM, ESS, and SM, clearly outperform CCDM. 
The more efficient signal space usage of ESS and SM becomes particularly apparent at very short blocklengths\footnote{The small differences in rate loss of ESS and SM for very short blocklengths result from the fact that they employ different orderings, and thus, the sets of sequences that they omit due to \eqref{eq:inputlength} are not identical.}. 
The inset of Fig.~\ref{rlosses} (left) shows the rate losses at $n=216$.
Here CCDM has 0.035 bits larger rate loss than MPDM.

\subsection{Achievable Information Rates}\label{ssec:air_results}
Here, we numerically study the AIRs of ESS, MPDM and CCDM in the finite blocklength regime.
As the figure of merit, the finite blocklength AIR for BMD is used as defined in~\cite[eq. (15)]{Fehenberger2019TCOM_MPDM}:
\begin{align}
    \text{AIR}_{n} = \rbmd-\rloss. \label{eq:AIRfl}
\end{align}
Here $\rbmd$ and $\rloss$ are as defined in \eqref{eq:rbmd} and \eqref{eq:rateloss}.
We note that \eqref{eq:AIRfl} converges to \eqref{eq:rbmd} when $n\rightarrow\infty$.
The finite-length AIR in \eqref{eq:AIRfl} has been employed to compare ESS and CCDM for the optical fibre channel in~\cite{Amari2019_IntroducingESSoptics,Amari2019_ESSreachincrease,Goossens2019_FirstExperimentESS}.
We note here that \eqref{eq:AIRfl} is an instance of the rate expression~\cite[eq. (1)]{BochererSS2019_PSandFECforFOComm} provided for the {\it layered PS} architecture\footnote{We refer the reader to~\cite[footnote 3]{Amari2019_IntroducingESSoptics} for a discussion on the derivation of \eqref{eq:AIRfl} from~\cite[eq. (1)]{BochererSS2019_PSandFECforFOComm}}.

In Fig.~\ref{rlosses} (right), $\text{AIR}_{n}$ in bit/1-D is shown versus SNR in dB for 8-ASK with ESS, MPDM and CCDM.
We use shaping blocks of length $n=216$, which is compatible to the $n_c=648$-bit LDPC codes of IEEE 802.11~\cite{80211-2016} that will be employed in PAS in subsequent sections.
Shaping algorithms operate at a rate of $k/n=1.75$, i.e., $k$ is set to 378 bits.
We note that this means we plotted the curves for fixed distributions and did not optimize them at each SNR, unlike~\cite[Fig. 4]{bocherer2015} or~\cite[Fig. 5]{Fehenberger2019TCOM_MPDM}.
For comparison, the Shannnon capacity $\frac{1}{2}\log(1+\snr)$ and the GMI for uniform 8-ASK are also plotted.
We observe that ESS and MPDM close most of the shaping gap.
From the inset figure, we see that ESS and MPDM are roughly 0.72 dB more SNR-efficient than uniform signaling at rate $R=2.25$.
We note that $R=2.25$ corresponds to $\gamma =R-k/n=0.5$, and thus, $R_c=5/6$ in the PAS context.
As a reference, the maximum possible capacity gain due to shaping at this rate is 1.04 dB.
The remaining gap of 0.32 dB is due to the finite blocklength nature of shaping and the discrete nature of the employed constellation.

From the inset of Fig.~\ref{rlosses} (right), we see that MPDM is 0.23 dB more power-efficient than CCDM.
This difference is consistent with the empirical relation between the power loss and rate loss explained in footnote 7, more specifically, $P_{\text{loss,dB}} \approx 7\rloss = 6\cdot  0.0352 = 0.21$~dB.

We conclude from Fig.~\ref{rlosses} that from a practical point of view, MPDM and \spsh~perform almost the same at blocklengths larger than $n\approx200$.
Therefore, to make a choice among these at such values of $n$, required storage, computational complexity and latency of the algorithms that can be used to implement MPDM and \spsh~should be considered.
We will discuss these aspects of shaping algorithms in Sec.~\ref{sec:complexity}.

\begin{remark}[{\bf Targeting a rate with DM}]\label{rmk:targetRforDM}
Example~\ref{ex:RLossMethod} shows that when the entropy of the target distribution is taken to be the target rate $k/n$ (1.75 in Example~\ref{ex:RLossMethod}), CCDM and MPDM are not able to obtain $2^k$ sequences.
This is due to the inevitable nonzero rate loss of the DM schemes for finite blocklengths.
For such cases, we increase the SNR that the target distribution is optimized for, until we obtain $2^k$ output sequences for the DM schemes.
\end{remark}

\subsection{End-to-End Decoding Performance}\label{end2end}
 In the following, the decoding performance is evaluated after transmission of 64-QAM over an AWGN channel.
  BRGC in Fig.~\ref{fig:PAShighlevel} (right) is used for amplitude to bit mapping after shaping, and for symbol mapping after FEC encoding as shown in Fig~\ref{fig:PAShighlevel} (left).
  Different transmission rates and length regimes of LDPC codes are considered. For each SNR, the simulations are run until at least 100 frame errors are observed. For the first case of long FEC we use codes from the DVB-S2 LDPC standard \cite{DVBS2} with blocklength $n_c=64800$~bits. In the case of short FEC the LDPC codes from the 802.11 standard~\cite{80211-2016} of length $n_c=648$~bits are used.
  
  For a fixed 1-D constellation size $M=2^m$, FEC code rate $R_c$ and target transmission rate $R$, we compute $\gamma=R_cm-(m-1)$ and accordingly, $k/n=R-\gamma$.
  Here the total number of 1-D symbols in an $n_c$-bit FEC codeword is $n=n_c/m$.
  For DM algorithms working with $\calA=\{1,3,5,7\}$, the AWGN-optimal MB PMFs at 10.7 and 14 dB SNR are quantized to obtain the integer composition based on~\cite{Bocherer2016TransIT_OptimumQuantize} for the target rates 4 and 4.5 bit/2-D, respectively.
  For \spsh~algorithms, $E^\bullet$ is selected as the minimum value that leads to $\Rs \geq k/n$.
  Both ESS and SM are then implemented with FP.
  The amplitude shaping function of SM is implemented using~\cite[Algorithm 1]{laroia1994}.

\begin{figure*}[t]
\centering
\includegraphics[width=\columnwidth]{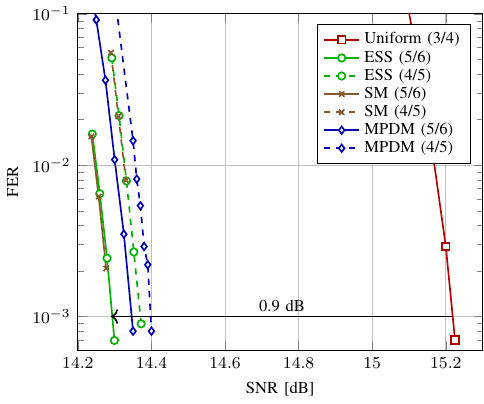}
\includegraphics[width=\columnwidth]{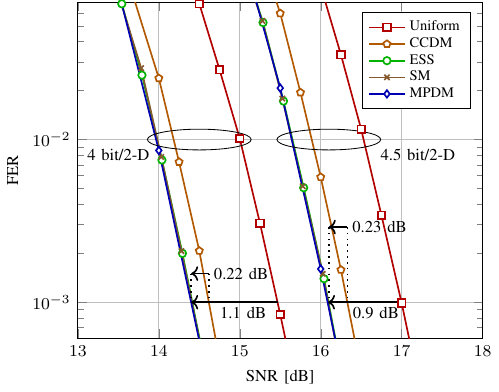}
\caption{(Left) FER vs. SNR for 64-QAM at a transmission rate of 4.5 bit/2-D. DVB-S2 LDPC codes with $n_c=64800$ bits are used. All shaping schemes use a blocklength of $n=180$. (Right) FER vs. SNR for 64-QAM at transmission rates of 4 and 4.5 bit/2-D. LDPC codes of 802.11 with $n_c=648$ bits are used. All shaping schemes use a blocklength of $n=216$.}
\label{FERcurves4halfDVBS2}
\end{figure*}

  Figure~\ref{FERcurves4halfDVBS2} (left) shows the decoding performance with DVB-S2 LDPC codes for ESS, SM, MPDM, and uniform signaling at a transmission rate of 4.5 bits per complex channel use (bit/2-D). 
  ESS, SM and MPDM, all of length 180 amplitudes, use either the LDPC code of rate 5/6 (solid curves) or rate 4/5 (dashed lines). 
  At this shaping blocklength, each LDPC frame consists of 120 shaped blocks. 
  In order achieve a transmission rate of 4.5 bit/2-D, the redundancy added by the shaping scheme is varied. 
  For uniform 64-QAM, the code rate is set to 3/4. 
  We observe for shaped schemes that the performance with FEC rate 5/6 is superior to rate 4/5, for which the reasons are outlined in Sec.~\ref{ssec:PASparameterselection}, and focus on 5/6 in the following. 
  
  At a FER of 1e-3, the shaped schemes outperform uniform signaling by approximately 0.9~dB. 
  We further note that ESS, SM and MPDM have very similar performance, with ESS and SM being approximately 0.05~dB more power-efficient than MPDM. This is in good agreement with the rate loss analysis of Fig.~\ref{rlosses} (left) where also only a marginal improvement of the \spsh~schemes over MPDM is found.
  
\begin{remark}
From the discussion in Sec.~\ref{ssec:PASparameterselection}, we expect an SNR improvement of approximately 0.83~dB of the shaped schemes over uniform signaling, which is in good agreement with the observed improvement of 0.9~dB. 
Potential reasons for the 0.1 dB difference between the theoretical analysis and the numerical simulations are the different coding gaps of the employed LDPC codes as well as the finite-length rate loss of the shaping schemes. 
\end{remark}
  
For short LDPC codes with shaped signaling, the shaping blocklength is set to $n=216$, which, in combination with the LDPC code length of $n_c=648$ bits and 64-QAM, gives a one-to-one correspondence between the blocklengths of FEC and shaping. In Fig.~\ref{FERcurves4halfDVBS2} (right), the decoding performance is analyzed at transmission rates of 4 and 4.5~bit/2-D. 
Uniform 64-QAM requires LDPC rates 3/4 and 5/6, respectively. 
For the shaped schemes, the code rate that minimizes $\delsnr$ for 64-QAM, and rates 4 and 4.5 bit/2-D can be computed to be $R_c \approx 0.79$ and 0.83 using \eqref{gapeq}, respectively.
Thus $R_c=5/6$ being the closest available to these values is used for shaped signaling. 

As shown in Fig.~\ref{FERcurves4halfDVBS2} (right), we observe that at rate 4~bit/2-D ESS, SM and MPDM, which have identical decoding performance in this setup, require 1.1~dB less SNR than uniform to achieve a FER of 1e-3. 
This improvement is due to the finite-length shaping gain as well as the reduced coding gap of the rate-5/6 LDPC code over the rate-3/4. 
We further observe that ESS and MPDM are 0.22~dB more power-efficient than CCDM.

Figure~\ref{FERcurves4halfDVBS2} (right) also shows the FER at rate 4.5~bit/2-D. 
Here, ESS, SM and MPDM again perform identically.
Uniform signaling is significantly outperformed by approximately 0.9~dB SNR. CCDM is now 0.23 dB less SNR-efficient than the other shaping approaches.

We have seen that the performance of the MPDM and \spsh~schemes is almost identical for the considered shaping length. 
Hence, implementation aspects, which are discussed next, are believed to be of significant importance in the comparison between these schemes.

\section{Approximate Complexity Discussion}\label{sec:complexity}
 In the preceding section, we followed the conventional approach of comparing different schemes by studying the blocklength that is required to obtain a certain shaping gain. 
 While this is certainly a natural choice for analysing and comparing shaped systems, this approach inherently assumes that shorter blocks are always better, for instance because they have advantages regarding implementation. 
 In the following, we comment on the implementation aspect by considering computational complexity, latency, and storage requirements.

 An example where slightly longer blocklengths can be beneficial also from an implementation perspective is the parallel-amplitude architecture proposed in \cite[Sec.~III]{Fehenberger2019Arxiv_PASR}. 
 By allowing a small additional rate loss, the throughput is increased significantly by using $\namp-1$ DMs in parallel. Furthermore, the serialism (and thus, the latency) of the SR method of \cite[Sec.~IV]{Fehenberger2019Arxiv_PASR} is smaller than AC-CCDM. It can thus be beneficial to make the blocks slightly larger than for conventional CCDM in order to facilitate implementation. 
 
 An interesting example where the selection of the shaping blocklength does not depend only on the complexity vs. shaping gain tradeoff is the nonlinear regime of the optical fibres. The authors of~\cite{Amari2019_IntroducingESSoptics} recently found that shaping over shorter blocklengths increases the nonlinear tolerance, and thus, the effective SNR.
 Their claim is that when the complexity considerations are ignored, there is an optimum $n$ that optimizes the balance between linear shaping gain and nonlinear tolerance.

\subsection{Latency}
In order to evaluate the latency of the discussed amplitude shaping  algorithms, we use the concepts of ``degree of serialism" and ``parallelization factor" as defined in~\cite{Fehenberger2019Arxiv_PASR}.
Degree of serialism is the number of loop iterations that are completed for shaping/deshaping operations.
We stress that this quantity neglects the computational complexity of these iterations, and thus the latency of the operations within each sequential processing step.
Therefore the degree of serialism can only serve as a rough indicator for latency.
On the other hand parallelization factor is the number of simultaneously possible executions of a process to complete shaping/deshaping operations.

AC, which can be employed to realize CCDM, is by nature a highly serial algorithm, and AC-CCDM has a serialism of $k$ for matching and $n$ for dematching~\cite{ccdm}.
SR-DM, which is an alternative to AC-CCDM in the binary-output case~\cite{Fehenberger2019Arxiv_PASR}, has a serialism of $\min(n_1,n-n_1)$ and 1 for shaping and deshaping, respectively\footnote{In the SR-DM context, $[n_1,n_2]$ is the composition of binary sequences at the output of the matcher.}.

In BL-DM~\cite{pikus2017} and PDM~\cite{SteinerSB2018_PDM}, a binary-output matcher is used for each of the $\log_2\namp=m-1$ amplitude bit levels to enable parallelization, and thus, the parallelization factor is $\log_2\namp$. 
As another attempt, PA-DM uses a binary-output matcher for $\namp-1$ of the $\namp$ amplitudes~\cite{Fehenberger2019Arxiv_PASR}, and thus, the parallelization factor is $\namp-1$.
A more detailed discussion on improving the parallelization of DM algorithms is provided in Sec.~\ref{ssec:distribution_matchers}.

The shaping and deshaping algorithms of ESS~\cite{willems1993} and~\cite[Algorithm 1]{laroia1994} have a serialism of $k$ and $n$, respectively. 
On the other hand SM~\cite{laroia1994} operates based on the D\&C principle as in~\cite{LangL1989_LeechLatticeModem}, and therefore has a serialism of $\log_2n$ for deshaping.
Table~\ref{tab:complexity} summarizes the serialism of discussed shaping schemes.

\subsection{Storage Requirements} \label{ssec:storage}
AC-CCDM, which employs an extension of~\cite{ramabadran1990} to nonbinary-output, associates an interval in $[0, 1)$ to each binary input sequence and to each constant composition amplitude sequence~\cite[Sec. IV]{ccdm}.
In simplified terms, the final interval is computed by recursively splitting the initial interval into $\namp$ subintervals.
The algorithm only requires the storage of the interval and the source statistics (i.e., the composition) which can be realized with $\log n$~bits\footnote{Here we assume that the memory required to store the interval is negligible, and roughly $\log_2 n$~bits are enough to store the composition which consists of numbers that add up to $n$.}.
Thus we denote the storage complexity of AC-CCDM by $\mathcal{O}(\log n)$.
A similar reasoning can be used to determine the storage complexity of SR-DM~\cite[Sec. IV]{Fehenberger2019Arxiv_PASR} which is also $\mathcal{O}(\log n)$.

\begin{table*}[t]
\renewcommand{\arraystretch}{1.2}
\centering
\caption{Serialism, required storage and computational complexity}
\resizebox{2\columnwidth}{!}{
\begin{tabular}{c|c|c|c|c||c|c|c|c|} 
\cline{2-9}
& \multicolumn{4}{c|}{\begin{tabular}{@{}c@{}} Direct Method \\ (Distribution Matching)\end{tabular}} & \multicolumn{4}{c|}{\begin{tabular}{@{}c@{}} Indirect Method \\ (Energy-Efficient Signal Space)\end{tabular}} \\\cline{2-9}
 & \multicolumn{2}{c|}{AC-CCDM~\cite{ccdm}} & \multicolumn{2}{c|}{SR-DM~\cite{Fehenberger2019Arxiv_PASR}} &  \multicolumn{2}{c|}{\begin{tabular}{@{}c@{}}ESS~\cite{willems1993} \\ \cite[Algorithm 1]{laroia1994}\end{tabular}} & \multicolumn{2}{c|}{SM~\cite{laroia1994}} \\
 \hline
 \multicolumn{1}{|c|}{\begin{tabular}{@{}c@{}} Serialism \\ (no. of loop iter.)\end{tabular}} & \multicolumn{2}{c|}{$k + n$} & \multicolumn{2}{c|}{$\min(n_1,n-n_1) + 1$} &  \multicolumn{2}{c|}{$k + n$} & \multicolumn{2}{c|}{$k + \log_2n$} \\
 \hline
 \multicolumn{1}{|c|}{\begin{tabular}{@{}c@{}} Storage \\ Complexity\end{tabular}} & \multicolumn{2}{c|}{$\mathcal{O}(\log n)$} & \multicolumn{2}{c|}{$\mathcal{O}(\log n)$} & \multicolumn{2}{c|}{\begin{tabular}{@{}c@{}}FP: $\mathcal{O}(n^3)$ \\ BP~\cite{GultekinWillems2018ISIT_EnumerativeShaping}: $\mathcal{O}(n^2\log n)$\end{tabular}} 
 & 
 \multicolumn{2}{c|}{\begin{tabular}{@{}c@{}}FP: $\mathcal{O}(n^2\log n)$ \\ BP~\cite{GultekinWillems2018ISIT_EnumerativeShaping}: $\mathcal{O}(n\log^2 n)$\end{tabular}} \\
 \hline
 \multicolumn{1}{|c|}{\begin{tabular}{@{}c@{}} Computations \\ (per 1-D)\end{tabular}} & \multicolumn{2}{c|}{\begin{tabular}{@{}c@{}} $n_a$ divisions, \\ multiplications \\ and comparisons \end{tabular}} & \multicolumn{2}{c|}{\begin{tabular}{@{}c@{}} Sh: $(n_a-1)$ BCs \\ Dsh: $(n_a-1)/2$ BCs \end{tabular}} & \multicolumn{2}{c|}{\begin{tabular}{@{}c@{}}Sh: $n_a$ comparisons and subtractions \\ Dsh: $n_a$ additions \\ (and $L$ comparisons/additions \\ per $n$-D for~\cite[Algorithm 1]{laroia1994}) \end{tabular}} 
 & 
 \multicolumn{2}{c|}{\begin{tabular}{@{}c@{}} Sh: $L$ multiplications, comparisons \\ and subtractions$^{\dagger}$ \\ Dsh: $L$ multiplications and additions \end{tabular}} \\
 \hline
 \multicolumn{9}{c}{\begin{tabular}{@{}c@{}} $^{\dagger}$SM requires a division per dimension for shaping as well. (Sh:Shaping, Dsh: Deshaping, FP: Full-precision, BP: Bounded-precision~\cite{GultekinWillems2018ISIT_EnumerativeShaping}, BC: Binomial Coefficient.)\end{tabular}
  }
\end{tabular}
}
\label{tab:complexity}
\end{table*}

In MPDM, in addition to the requirements of the underlying CCDM algorithm, a composition is chosen based on a prefix of the binary input sequence.
For this purpose, a prefix code and the corresponding Huffman tree is constructed~\cite[Sec. III-C]{Fehenberger2019TCOM_MPDM}.
To store the binary-tree, a LUT can be constructed.
The size of this table depends on the number of utilized compositions and grows with $n$ for a fixed $\calA$ and $k/n$.
For practical scenarios, the number of compositions is on the order of a few hundreds as shown in the following example.

\begin{example}[{\bf MPDM, number of compositions}]
We consider $\calA = \{1, 3, 5, 7\}$, $n=216$ and target rates $k/n = 1.5$ and 1.75~bit/1-D.
To obtain these target rates, MPDM uses 318 and 593 different compositions, respectively.
Note that these are the parameters that are used for the simulations considered in Fig.~\ref{FERcurves4halfDVBS2} (right).
\end{example}

FP implementations of ESS and~\cite[Algorithm 1]{laroia1994} require the storage of an $n$-by-$L$ matrix where each element is at most $\lc n\Rs \rc$-bits long.
Thus following Remark~\ref{rem:L_func_N}, the storage complexity of these algorithms is $\mathcal{O}(n^3)$ for fixed $\Rs$.
FP SM can be realized by storing a $\log_2 n$-by-$L$ matrix~\cite{laroia1994}, which has complexity $\mathcal{O}(n^2\log n)$.
We note here that these values are in alignment with~\cite[Table I]{laroia1994}. 

\begin{example}[{\bf FP \spsh, required storage}]
To realize ESS or~\cite[Algorithm 1]{laroia1994} for the setup in Example~\ref{ex:sphereshaping}, at most $Ln\lc n\Rs \rc = 80.46$~kilobytes (kB) of memory is required.
On the other hand for SM, at most $L\log_2n\lc n\Rs \rc=7.54$~kB of memory should be allocated.
\end{example}

\begin{remark}\label{rem:mant_e}
To compute the required storage for \spsh~in the BP case, we will assume that $n_m$ is independent of $n$.
This assumption relies on the fact that the rate loss resulting from BP only depends on $n_m$~\cite{GultekinWillems2018ISIT_EnumerativeShaping}.
Thus for a fixed rate loss, the required value of $n_m$ is independent of $n$.
Expressing the number of bits to store the exponent as $n_p = \lc \log_2 \left( \lc n\Rs \rc - n_m \right) \rc$, we see that $n_p$ behaves as $\log_2 n$ for a fixed $n_m$.
We note here that for a fixed $n$, $\calA$ and target $k$, the natural choice for $n_m$ is the smallest value that keeps the number of sequences at least $2^k$~\cite{GultekinWillems2018ISIT_EnumerativeShaping}.
\end{remark}

For the BP implementations of ESS, SM and~\cite[Algorithm 1]{laroia1994}, each element of the stored shaping matrix is at most $(n_m+n_p)$-bit long~\cite{GultekinWillems2018ISIT_EnumerativeShaping}.
Following Remark~\ref{rem:mant_e}, the storage complexity of ESS and~\cite[Algorithm 1]{laroia1994} in the BP case is $\mathcal{O}(n^2\log n)$.
On the other hand the storage complexity of BP SM is $\mathcal{O}(n\log^2 n)$.

\begin{example}[{\bf BP \spsh, required storage}]
To realize ESS or~\cite[Algorithm 1]{laroia1994} with $n_m=9$ and $n_p=7$ for the setup in Example~\ref{ex:sphereshaping}, at most $Ln(n_m+n_p)=11.39$~kB of memory is required.
On the other hand, when implemented using $n_m=6$ and $n_p=7$, SM demands at most $L\log_2n(n_m+n_p) = 0.87$~kB of memory.
We note that the mantissa lengths $n_m$ are selected according to the discussion in Remark~\ref{rem:mant_e}.
\end{example}

In conclusion, we believe that storage requirements in the order of a few kB are not critical for high-throughput operation, particularly in comparison to latency and complexity.
Note that the required storage for BL-DM, PDM and PA-DM depends on the underlying algorithm.

\subsection{Computational Complexity}
To comment on the computational complexity of the amplitude shaping algorithms, we will mainly consider the number of required bit operations or computations of binomial coefficients (BC).
The caveat here is that this approach only gives a rough estimate since the complexity of an operation depends heavily on the specific case that it is executed in.
As an example, the seemingly simple operation of comparing the sizes of two numbers can be computationally challenging for large numbers.
On the other hand the notoriously expensive division operation reduces to a simple shift in registers for some specific divisors.

As explained in Sec.~\ref{ssec:storage}, AC-CCDM can be realized by splitting an interval into $\namp$ per 1-D.
This requires at most $\namp$ multiplications.
For each multiplication, one of the multipliers is found by a division using the statistics of the composition.
Finally, at most $\namp$ comparisons are carried out.
We note that practical discussions such as ``numerical precision", ``gaps between intervals" and ``rescaling" are omitted here, and the reader is referred to~\cite{RissanenL1979_AC,Langdon1984_Intro2AC,sayood2002lossless} for details.

An approximate implementation of AC-CCDM is proposed in~\cite{streamingdm} where computations are realized with fixed-point operations.
However, this implementation also requires multiplications, divisions and comparisons of large integer numbers.
In addition, an implementation of AC-DMs based on finite-precision arithmetic is provided in~\cite{PikusXK2019_FPACDM}.

SR-CCDM, in contrast to AC, is based on calculating binomial coefficients (BCs). 
Thus, the number of bit operations depends a lot on how this computation is implemented or whether the BCs can be pre-computed and stored.
However, to give a rough indication of computation complexity, we need to compute $(\namp-1)$ BCs for unranking (shaping), and $(\namp-1)/2$ BCs for ranking (deshaping) per 1-D.

When ESS and~\cite[Algorithm 1]{laroia1994} are implemented with FP, at most $\namp$ additions (subtractions) of numbers from the corresponding shaping matrix are required per 1-D.
These numbers are at most $\lc n\Rs \rc$-bit long, i.e., $\namp\lc n\Rs \rc$ bit operations\footnote{``Bit operation" refers to one-bit addition or subtraction.} per dimension (bit oper./1-D) are necessary.
Thus, the computational complexity of these algorithms is $\mathcal{O}(n)$.
FP implementation of SM however, requires at most $L$ multiplications of numbers from the shaping matrix.
Therefore, the computational complexity of SM is $\mathcal{O}(n^3)$.\footnote{Multiplication of two $k$-bit numbers is assumed to be equivalent to $k^2$~bit operations following Laroia {\it et al.}~\cite{laroia1994}.}

\begin{example}[{\bf FP \spsh, computational complexity}]
Based on Example~\ref{ex:sphereshaping}, at most $\namp\lc n\Rs \rc=452$~bit oper./1-D are necessary to realize ESS and~\cite[Algorithm 1]{laroia1994}.
On the contrary, for SM algorithms, at most $L\lc n\Rs \rc^2=1136441$~bit oper./1-D are required.
\end{example}

With BP approach, ESS and~\cite[Algorithm 1]{laroia1994} can be implemented with at most $\namp(n_m+n_p)$~bit oper./1-D.
Then their computational complexity is $\mathcal{O}(\log n)$.
On the other side, BP SM can be realized with at most $L(n_m+n_p)^2$~bit oper./1-D.
Therefore the complexity of SM is now $\mathcal{O}(n \log^2 n)$.

\begin{example}[{\bf BP \spsh, computational complexity}]
When Example~\ref{ex:sphereshaping} is now constructed with $n_m=9$ and $n_p = 7$, ESS and~\cite[Algorithm 1]{laroia1994} require at most $\namp(n_m+n_p)=64$~bit oper./1-D.
Correspondingly, if SM is realized with $n_m=6$ and $n_p=7$, $L(n_m+n_p)^2 = 15041$~bit oper./1-D are necessary.
\end{example}

Table~\ref{tab:complexity} summarizes serialism, required storage and computational complexity of discussed shaping algorithms as classified in Fig.~\ref{fig:shapingtaxonomy}.
The main conclusion from Table~\ref{tab:complexity} is that for DM, AC and SR provide a tradeoff between serialism and computational complexity. 
However, we note that SR can only be used for binary-output DM.
On the other hand for \spsh, SM and ESS create a tradoff between required storage and computational complexity.
The selection among different algorithms then depends on the actual resources that are available for shaping in practice, and thus, we refrain from making definitive suggestions here.

\begin{figure}[t]
\centering
\includegraphics[width=\columnwidth]{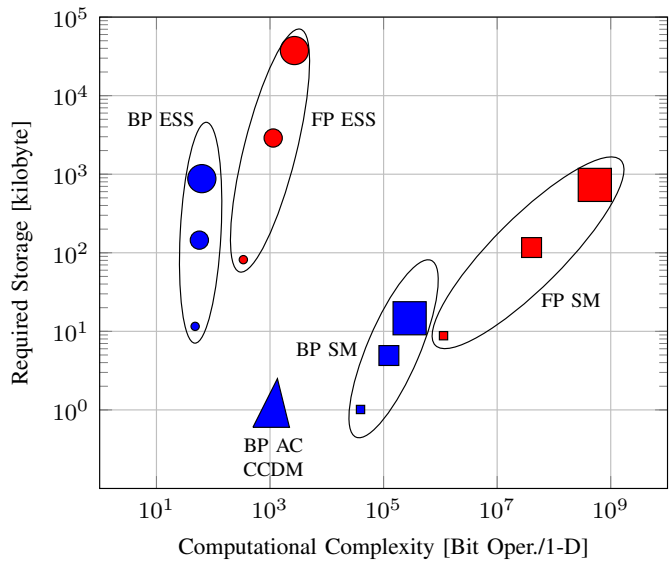}
\caption{Maximum computational complexity vs. maximum required storage of ESS and SM.
\textcolor{red}{Red}- and \textcolor{blue}{blue}-colored markers indicate FP and BP implementations, respectively. 
Radii of the markers are proportional to the corresponding blocklength $n \in \{64, 216, 512\}$.
Here we assume that BP AC-CCDM is implemented with finite-precision arithmetic using 16-bit numbers which is comparable to the values selected in~\cite{PikusXK2019_FPACDM}.}
\label{fig:complexfigure}
\end{figure}

We conclude this paper by showing in Fig.~\ref{fig:complexfigure}, the maximum required storage versus maximum number of computations required to implement BP and FP \spsh, and BP AC-CCDM\footnote{BP AC-CCDM refers to the finite-precision implementation of AC-CCDM as discussed in~\cite{PikusXK2019_FPACDM}.}.
We see that there is a computational complexity vs. required storage tradeoff between ESS (and~\cite[Algorithm 1]{laroia1994}) and SM.
ESS requires larger storage but can be implemented with a smaller complexity, and only demands additions and subtractions.
On the other hand, SM can be realized with a smaller storage, however requires many multiplications and divisions.
In fact, by modifying the corresponding shaping and deshaping algorithms, it is also possible to adjust the balance between computational complexity and required storage as explained in~\cite[Sec. 4.3.4]{Fischer2002_PrecodingShaping}, i.e., operate between the ESS and SM clusters in Fig.~\ref{fig:complexfigure}.
Furthermore, there is also a difference in computational complexities of ESS and~\cite[Algorithm 1]{laroia1994}.
An initial step is required in~\cite[Algorithm 1]{laroia1994} where the $n$-shell that the corresponding sequence is located on is determined.
This step requires at most $L-1$ additions and comparisons.

Finally, Fig.~\ref{fig:complexfigure} also shows that BP AC-CCDM can be implemented with moderate computational complexity and minimal storage. 
Furthermore, these requirements do not heavily depend on blocklength $n$.
Thus for large $n$ where its rate losses are small, and for applications for which high serialism of AC is not important, AC-CCDM is an effective and low-complexity choice as a shaping algorithm.

\section{Conclusion}\label{sec:conclusion}
This paper reviewed prominent amplitude shaping architectures and algorithms for the probabilistic amplitude shaping (PAS) framework.
Constant composition distribution matching (CCDM), multiset-partition DM (MPDM) and sphere shaping (\spsh) are all optimum shaping techniques for asymptotically large blocklengths, in the sense that they have vanishing rate loss.
However, for short blocklengths, CCDM addresses a smaller set of output sequences than that of MPDM and \spsh, leading to higher rate losses.
We provided evidence for the AWGN channel that seeking to utilize the signal space in energy-efficient manners is better than attempting to obtain the capacity-achieving distribution, which is derived for asymptotically large, and thus, impractical blocklengths.
Therefore, MPDM, \spsh, and other energy-efficient shaping architectures are suitable to be used over a wider blocklength regime, especially for blocklengths below a couple of hundred symbols.

In addition to the rate loss analysis, we evaluated the achievable information rates (AIR) and frame error rates (FER) of the PAS framework employing CCDM, MPDM and \spsh~as the underlying amplitude shaping approach.
Enumerative sphere shaping (ESS) and shell mapping (SM) are both considered as potential \spsh~implementations.
AWGN channel simulations with 64-QAM demonstrate that power-efficiency gains on the order of 1 dB can be obtained already at blocklengths around 200 by employing MPDM and \spsh, and thus, justify our earlier observation on the objective of amplitude shaping.
CCDM provides gains around 0.75 dB for the same settings.
Furthermore, these gains are predicted well by shaping gain and AIR computations based on bit-metric decoding.

In the last part of the paper, we discussed the performance of shaping algorithms considering latency, required storage and computational complexity.
To realize DM, arithmetic coding (AC)-based implementation of MPDM requires minimal storage and can be implemented with a few computations per input symbol.
However AC has a higher serialism than subset ranking (SR)-based implementation which on the other hand has increased computational complexity.
For \spsh, ESS and SM provide a tradeoff between storage and computational complexities, where the complexity is more due to the required storage for ESS and required number of computations for SM.
Thus the decision on which algorithm should be used to realize energy-efficient amplitude shaping depends on the application-specific requirements on latency, available storage and tolerable computational complexity.

\end{document}